\documentclass[12pt,preprint,useAMS,usenatbib]{emulateapj}
\usepackage{graphicx, txfonts}
\usepackage{setspace}
\usepackage{natbib}
\usepackage{ulem}
\usepackage{url}
\usepackage{multirow}

\usepackage[usenames,dvipsnames]{color}
\newcommand{\CII}{C\,\textsc{ii}}

\newcommand{\DI}{\textrm{D}\,\textsc{i}}
\newcommand{\FeII}{Fe\,\textsc{ii}}

\newcommand{\HI}{\textrm{H}\,\textsc{i}}

\newcommand{\HII}{\textrm{H}\,\textsc{ii}}

\newcommand{\Lya}{Ly$\alpha$}
\newcommand{\lya}{Ly$\alpha$}

\newcommand{\NHI}{$N(\textrm{H}\,\textsc{i})$}

\newcommand{\NI}{N\,\textsc{i}}

\newcommand{\OI}{O\,\textsc{i}}

\newcommand{\SiII}{Si\,\textsc{ii}}
\newcommand{\SiIII}{Si\,\textsc{iii}}

\newcommand{\yp}{${Y}_{\rm P}$}
\newcommand{\dhp}{(D\,/\,H)$_{\rm p}$}
\newcommand{\lih}{$^7$Li/H}
\newcommand{\neff}{$N_{\rm eff}$}
\newcommand{\Obary}{$\Omega_{\rm b,0}$}
\newcommand{\hsq}{$\,h^2$}
\newcommand{\dHe}{$d$($p,\gamma$)$^3$He}
\newcommand{\alis}{\textsc{alis}}

\shorttitle{Precision measures of the primordial abundance of deuterium}
\shortauthors{Cooke et al.}

\begin{document}

\title{Precision measures of the primordial abundance of deuterium\altaffilmark{$\star$}}

\author{Ryan J. Cooke\altaffilmark{1,7}, Max Pettini\altaffilmark{2,3}, Regina A. Jorgenson\altaffilmark{4}, Michael T. Murphy\altaffilmark{5}, Charles C. Steidel\altaffilmark{6}}

\altaffiltext{$\star$}{Based on observations collected at the European Organisation for Astronomical Research 
in the Southern Hemisphere, Chile [VLT program IDs:
68.B-0115(A), 70.A-0425(C), 078.A-0185(A), 085.A-0109(A)],
and at the W.M. Keck Observatory
which is operated as a scientific partnership among the California Institute of 
Technology, the University of California and the National Aeronautics and Space 
Administration. The Observatory was made possible by the generous financial
support of the W.M. Keck Foundation.
Keck telescope time was partially granted by NOAO,
through the Telescope System Instrumentation Program
(TSIP). TSIP is funded by NSF.}
\altaffiltext{1}{Department of Astronomy and Astrophysics, UCO/Lick Observatory, University of California, Santa Cruz, CA 95064, USA}
\altaffiltext{2}{Institute of Astronomy, Madingley Road, Cambridge, UK, CB3 0HA}
\altaffiltext{3}{Kavli Institute for Cosmology, Madingley Road, Cambridge, UK, CB3 0HA}
\altaffiltext{4}{Institute for Astronomy, University of Hawaii, 2680 Woodlawn Drive, Honolulu, HI 96822, USA}
\altaffiltext{5}{Centre for Astrophysics and Supercomputing, Swinburne University of Technology, Hawthorn, Victoria 3122, Australia}
\altaffiltext{6}{California Institute of Technology, MS 249-17, Pasadena, CA 91125, USA}
\altaffiltext{7}{Morrison Fellow;~~~~email: rcooke@ucolick.org}
\date{\today}

\begin{abstract}
We report the discovery of deuterium absorption in the very metal-poor
([Fe/H]\,=$-2.88$) damped Lyman-$\alpha$ system at $z_{\rm abs} = 3.06726$
toward the QSO SDSS~J1358+6522.
 On the basis of 13 resolved \DI\ absorption
lines and the damping wings of the \HI\ Lyman $\alpha$ transition, 
we have obtained a new, precise measure of the primordial abundance 
of deuterium. Furthermore, to
bolster the present statistics of precision D/H measures, we have reanalyzed all of the
known deuterium absorption-line systems that satisfy a set of strict criteria.
We have adopted a blind analysis strategy (to remove human bias), and
developed a software package that is specifically designed for precision
D/H abundance measurements. For this reanalyzed sample of systems, we
obtain a weighted mean of \dhp\ $=(2.53\pm0.04)\times10^{-5}$,
corresponding to a universal baryon density 100\,\Obary\hsq$=2.202\pm0.046$
for the standard model of big bang nucleosynthesis (BBN). By combining our
measure of \dhp\ with observations of the cosmic microwave background (CMB),
we derive the effective number of light fermion species, \neff $= 3.28\pm0.28$.
We therefore rule out the existence of an additional (sterile) neutrino
(i.e. \neff $= 4.046$)  at 99.3 per cent confidence ($2.7\sigma$),
provided that the values of \neff\ and of the baryon-to-photon ratio ($\eta_{10}$) 
did not change between BBN and recombination.
We also place a strong bound on the
neutrino degeneracy parameter, independent of the $^4$He primordial
mass fraction, \yp : $\xi_{\rm D}=+0.05\pm0.13$ based
only on the CMB+\dhp\ observations.
Combining this value of $\xi_{\rm D}$ with the
current best literature measure of \yp, we find a $2\sigma$
upper bound on the neutrino degeneracy parameter, $|\xi|\le+0.062$.
\end{abstract}

\keywords{quasars: absorption lines -- quasars: individual: J1358+6522 -- cosmology: observations.}

%%%%%%%%%%%%%%%%%%%%%%
\section{Introduction}
%%%%%%%%%%%%%%%%%%%%%%

We are currently in an exciting era of high-precision cosmology,
with most of the ``standard'' cosmological model parameters
now known to within a few percent. In particular, the recent analysis
of the cosmic microwave background (CMB) temperature fluctuations
recorded by the \textit{Planck} satellite has found that the present day
density of baryons, \Obary, contributes just $(2.205\pm0.028)/$\hsq\
per cent of the critical density \citep{Efs13}, where $h$ is the Hubble constant
in units of 100 km s$^{-1}$ Mpc$^{-1}$. \textit{Planck}'s determination of \Obary\hsq,
which is now limited by cosmic variance, is the most precise measure of
this fundamental physical quantity for the foreseeable future
(when derived from the CMB).

For a long time (e.g. \citealt{WagFowHoy67}), it has been appreciated that a complementary measurement of
\Obary\hsq\ can be deduced from the relative abundances of the
light elements that were created during big bang nucleosynthesis (BBN). Aside
from protons, the only stable nuclei that were produced in astrophysically accessible
abundances are $^2$H (a.k.a. deuterium, D), $^3$He, $^4$He, and trace amounts
of $^7$Li (see \citealt{Ste07} for a comprehensive review). 
In particular, considerable effort
has been devoted to measuring the abundance of deuterium (D/H), the mass
fraction of $^4$He (\yp), and the abundance of $^7$Li. Of these, the primordial abundance of
deuterium is generally accepted as the best `baryometer', owing to its sensitivity and
monotonic relationship to \Obary\hsq.

The mass fraction of $^4$He, on the other hand, is relatively insensitive to
\Obary\hsq, but depends quite strongly on the expansion rate
of the early Universe (typically parameterized in non-standard models of
BBN as an effective number of neutrino species, \neff). At present,
this measurement is primarily limited by systematic uncertainties in
converting the He and H emission lines of metal-poor \HII\ regions
into an estimate of \yp\ (see e.g. \citealt{IzoStaGus13,Ave13}). Unlike \yp, the
\lih\ ratio depends modestly on both \Obary\hsq\ and \neff. However, the
observationally inferred value for the ``primordial'' level, derived from
metal-poor Galactic halo stars \citep{SpiSpi82,Mel10}, is discrepant
by a factor of $\sim3$ compared with the standard model predictions,
becoming even more discrepant at the lowest metallicities \citep{Sbo10}.

As pointed out recently by \citet{NolHol11} (see also \citealt{Cyb04}),
\textit{precise} measures of the primordial D/H ratio can provide interesting
bounds on \neff, when combined with a measure of \Obary\hsq\ from the CMB.
The promise of this method was recently demonstrated by \citet{PetCoo12},
who obtained the most precise measure of the D/H ratio to date, using a
metal-poor ([Fe/H]$=-2.33$) damped Lyman~$\alpha$ system (DLA), seen
in absorption against a bright background quasar.

The prospect of measuring the D/H ratio in gas clouds seen in absorption
along the line-of-sight to a high-redshift quasar was first pointed out by \citet{Ada76}.
This vision was only realized much later, with the advent of 8-10\,m class telescopes
equipped with high resolution echelle spectrographs \citep{BurTyt98a,BurTyt98b}.
Since these first discoveries, a handful of additional cases have been identified
(see \citealt{PetCoo12} for the `\textit{top-ten}' list), including one system that appears
to be chemically pristine \citep{FumOmePro11}. One lingering concern with this
prime set of measurements is that the dispersion in the reported D/H measures
is significantly larger  than the quoted errors (first noted by \citealt{Ste01}).
Indeed, a simple $\chi^{2}$ test reveals that the errors for all D/H measurements
would need to be scaled upwards by $\sim33\%$, if the observed dispersion were due
to chance alone. An alternative possibility, if the quoted random errors are in fact
realistic, is that the analyses may suffer from small systematic biases that
are now important to recognize and correct for.

%%%%%%%%%%%%
% FIGURE 1 %
%%%%%%%%%%%%
\begin{figure*}
  \centering
  \includegraphics[angle=0,width=17.0cm]{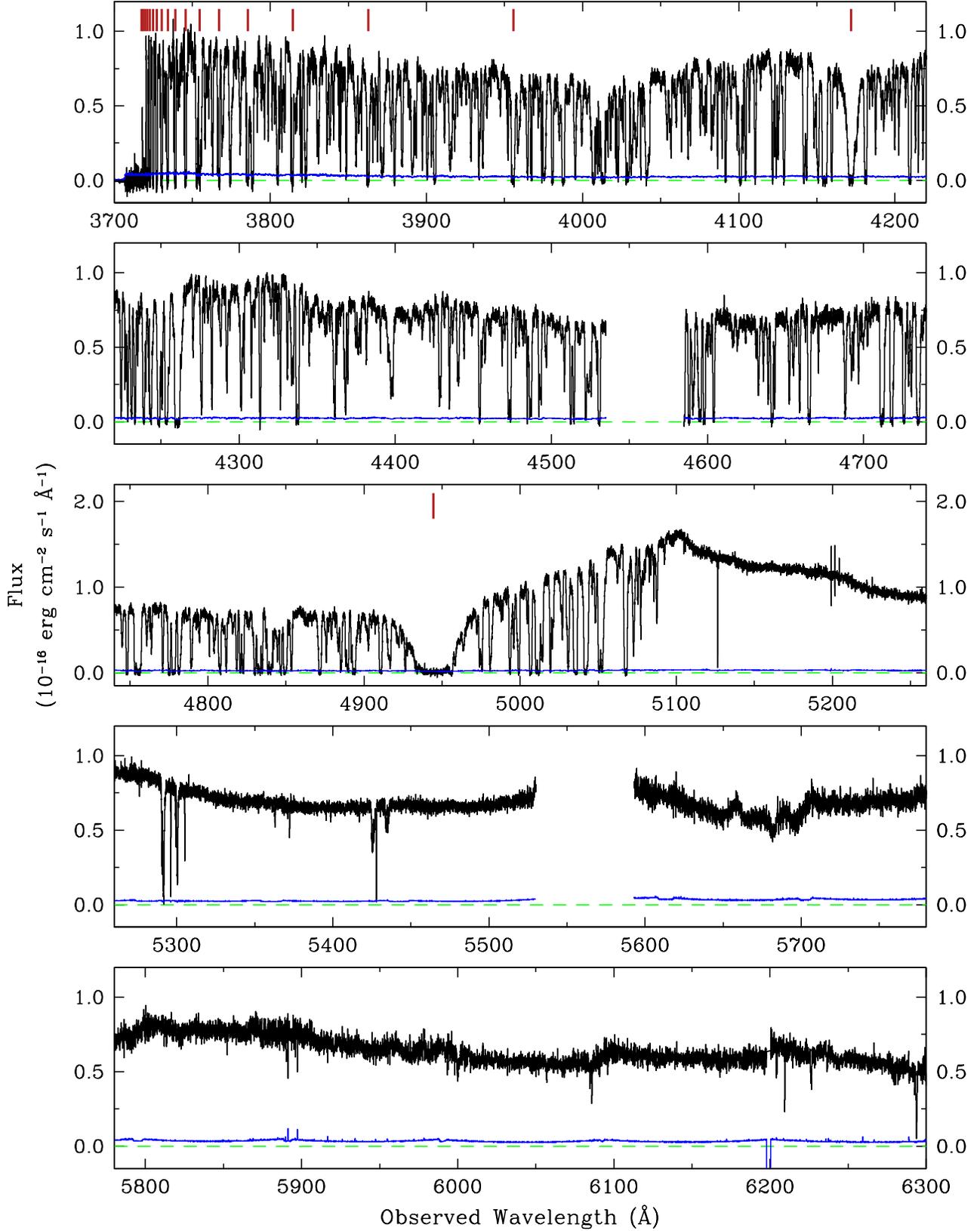}
  \caption{ 
Flux-calibrated spectrum of J1358$+$6522 (black histogram). The zero-level and
error spectrum are shown as the green dashed and solid blue lines respectively.
The \HI\ Lyman series lines of the DLA are indicated by red tick marks. Since the
SDSS data do not extend below 3800\,\AA, we extrapolated the response curve
for fluxing the data a further 100\,\AA\ to the blue.
  }
  \label{fig:fluxqso}
\end{figure*}

This concern has prompted us to identify the rare handful of systems which
afford \textit{precise} measurements of the primordial abundance of deuterium
together with realistic error estimates. This effort is part
of our ongoing survey to study
the chemistry of the most metal-poor
DLAs, described in more detail 
in \citet{Pet08a} and \citet{Coo11}.

In this paper, we present the \textit{Precision Sample} of
primordial deuterium abundance measurements, consisting
of the current best estimates of D/H in QSO
absorption line systems. We also report a DLA where a new,
precise measurement of the primordial deuterium abundance
could be obtained. All of these systems have been reanalyzed
in a self-consistent manner, taking into account the dominant
sources of systematic uncertainty. In the following section, we provide
the details of the observations and data reduction procedures for the
newly observed DLA, and discuss the selection criteria we have adopted
to define the Precision Sample. In Section~\ref{sec:profanalysis}, we
describe the analysis of the new system with deuterium absorption,
and estimate its D/H ratio. In Section~\ref{sec:cosmology} 
we derive the value of \dhp\ from the Precision Sample,
discuss its cosmological implications, and consider
current limitations in using D/H
abundance measurements for cosmology.
Section~\ref{sec:newphysics} deals with the
implications of our results for physics beyond
the standard model, considering in particular bounds 
on the effective
number of neutrino families and on
the lepton asymmetry.  
Finally, in Section~\ref{sec:conc}, we summarize the
main findings from this work.

\section{Observations and data reduction}
\label{sec:obs}

\subsection{The DLA towards J1358$+$6522}

Most newly discovered DLAs are now identified using the low-resolution spectra
provided by the Sloan Digital Sky Survey (SDSS). Recent searches for DLAs have
yielded several thousand systems \citep{ProWol09,Not12}.
At the modest resolution ($R\sim 2000$) and
signal-to-noise ratio (S/N$\,\sim20$) of the SDSS spectra, the metal absorption
lines of the most metal-poor DLAs are unresolved and undetected. Follow-up,
high spectral resolution observations are thus required to pin-down their chemical
abundances (see \citealt{Coo11}).

\citet{Pen10} were the first to recognize the very metal-poor DLA at redshift
$z_{\rm abs}\simeq3.0674$ towards the $z_{\rm em}\simeq3.173$ quasar
SDSS J1358$+$6522. On the basis of their intermediate resolution 
observations (60\,km s$^{-1}$ full-width at half maximum; FWHM) with the
Keck Echellette Spectrograph and Imager, these authors concluded
that this DLA has a metallicity of [Fe/H] $\le-3.02$, and is thus among the most 
metal-poor known \citep{Coo11}.
We re-observed this QSO with the Keck High Resolution Echelle
Spectrometer (HIRES; \citealt{Vog94}) using the red cross-disperser to
perform an accurate chemical abundance analysis \citep{Coo12}.
The HIRES data pinned down the metallicity of the DLA ([Fe/H]\,$ = -2.84\pm$0.03)
and provided the first indication for the presence of a handful of
resolved \DI\ absorption lines.

Taking advantage of the high efficiency of the HIRES UV cross-disperser
at blue wavelengths, we then performed dedicated HIRES observations
with the goal of obtaining a precise measure of the deuterium abundance.
Our observations included a total of $24\,300$\,s on 2012, May 11, (divided
equally into nine exposures) and an additional $29\,750$\,s on 2013, May 3 and 5.
An identical instrument setup was employed for both runs; we used the
C1 decker ($7''\times0.861''$, well-matched to the seeing conditions of $0.6''-0.8''$),
which delivers a nominal spectral resolution of
$R\sim48\,000$ ($\equiv6.3$ km s$^{-1}$ FWHM)
for a uniformly illuminated slit. Our setup covers the wavelength
range $3650-6305$\,\AA, with small gaps near 4550\,\AA\ and 5550\,\AA\
due to the gaps between the HIRES detector chips.
The exposures were binned on-chip $2\times2$.

The data were reduced following the standard procedures of bias subtraction,
flat-fielding, order definition and extraction. The data were mapped onto a
vacuum, heliocentric wavelength scale from observations of a ThAr lamp
which bracketed the science exposures. To test the quality of the data reduction,
we reduced the data using two pipelines: \textsc{makee} which is maintained by
T.~Barlow\footnote{\textsc{makee} is available
at:\\ \texttt{http://www.astro.caltech.edu/$\sim$tb/makee/}},
and \textsc{HIRedux} which is maintained by
J.~X.~Prochaska\footnote{\textsc{HIRedux} can be obtained
from:\\ \texttt{http://www.ucolick.org/$\sim$xavier/HIRedux/index.html}}.
We found that the skyline subtraction performed by \textsc{HIRedux} was superior
to that of \textsc{makee}, and substantially improved the S/N
of the spectra at blue wavelengths. This was very important for the removal of
telluric features near the redshifted \DI\ and \HI\ transitions of the DLA. All data
analyzed and presented herein for J1358$+$6522 were reduced with \textsc{HIRedux}.
The individual orders were combined with \textsc{UVES\_popler}, which is maintained
by M.~T.~Murphy\footnote{\textsc{UVES\_popler} can be downloaded
from:\\ \texttt{http://astronomy.swin.edu.au/$\sim$mmurphy/UVES\_popler/}}.
As a final step, we flux-calibrated the
echelle data by comparison to the SDSS data.
The combined S/N ratio of the final spectrum per 2.5 km s$^{-1}$ pixel
is $\sim30$ near 4000\,\AA, $\sim45$ at 5000\,\AA\ and $\sim20$ near 6000\,\AA.
The reduced, flux-calibrated
spectrum is presented in Figure~\ref{fig:fluxqso}, where the red tick marks
indicate the wavelengths of the redshifted \HI\ Lyman series transitions.

\subsection{Literature D/H QSO Absorption Line Systems}
\label{sec:litsyst}

\subsubsection{Selection Criteria}
\label{sec:criteria}

In this section, we outline the strict set of rules that we have
used to define the Precision Sample. Our goal is to identify
the small handful of systems currently known where the most 
accurate and precise measures of D/H can potentially be
obtained. The set of restrictions we applied are as follows:
\begin{itemize}
\item We require that the \HI\ \Lya\ absorption line must exhibit
Lorentzian damping wings. This criterion is satisfied for absorption
line systems with log \NHI/cm$^{-2}\, \ge19$. Such absorption
lines lie on the damping regime of the curve-of-growth,
where \NHI\ can be derived independently of the cloud model;
the damping wings uniquely determine the total \HI\ column density.
\item Due to the power of the \HI\ \Lya\ transition in establishing
the total \HI\ column density, we require that the wings of this
transition are not strongly blended with nearby, unrelated,
strong \HI\ absorption systems. If there exist additional
strong \HI\ absorption systems nearby, we impose
that the edges of their \Lya\ absorption troughs 
should be separated by at least 500 km s$^{-1}$,
and that all such absorption systems should be modelled simultaneously.
\item At least two, resolved and apparently unblended, optically
thin transitions of \DI\ must be available. This ensures that the
total \DI\ column density can be measured accurately, and
independently of the cloud model.
\item The data must have been obtained with a high resolution
echelle spectrograph, with
$R\ge30\,000$ (i.e. $v_{\rm FWHM} \le 10$ km s$^{-1}$), to
resolve the broadening of the \DI\ lines, and recorded at
S/N $\gtrsim 10$ per pixel ($\sim3$ km s$^{-1}$) at both
\Lya\ and the weakest \DI\ absorption line used in the analysis.
\item Several unblended metal lines with a range of oscillator
strengths for a given species (ideally \OI\ and \SiII) must be
present if there is only one optically thin \DI\ transition
in order to determine the velocity structure of the absorbing
cloud, and ensure that the presence of any partly ionized gas
(which may contribute to the \HI\ column density) is
accurately modelled. However, if there exist at least
two optically thin \DI\ transitions, metal lines are not strictly
required.
\end{itemize}

Of course, we are also limited by data that we, and others,
have access to. Data that are not publicly available on archives
could not be used in this study. In total, there are four systems in
the literature that meet the above criteria. We now give a 
brief description of these systems.

\subsubsection{HS0105$+$1619, $z_{\rm abs}=2.53651$}

Keck+HIRES observations of the QSO HS\,0105$+$1619
were obtained as follows:
1800\,s in 1999 (Program ID: U05H, PI: A.~Wolfe),
85\,000\,s in 1999--2000 by O'Meara et al. (2001),
and  21\,500\,s  in 2005 (Program ID: G10H, PI: D.~Tytler).
The data obtained by \citet{OMe01} are not available
from the Keck Observatory Archive.
Thus, we have reanalysed only the last dataset which,
in any case, was obtained with
a much improved HIRES detector.
As reported by \citet{OMe01},
the spectrum shows nine clean \HI\ transitions in the Lyman series
together with four optically thin \DI\ absorption lines;
the absorption is confined to
a single velocity component at $z_{\rm abs}=2.536509$.
The first ions exhibit slightly asymmetric profiles,
arising from ionized gas blueshifted by $7.1$ km s$^{-1}$
relative to the main component
(see Figure~\ref{fig:HS0105p1619}).
The \HI\ column density of this ionized
component contributes just 0.01 dex to the total column
density $\log N$(H\,{\textsc i})/cm$^{-2} = 19.42 \pm 0.01$.
The data and best-fitting
model are shown in Figure~\ref{fig:HS0105p1619}.

\subsubsection{Q0913$+$072, $z_{\rm abs}=2.61829$}

The quasar Q0913$+$072 intersects a DLA 
with $\log N$(H\,{\textsc i})/cm$^{-2} = 20.34 \pm 0.04$ at
$z_{\rm abs}=2.61829$. 
Q0913$+$072 was observed with the Ultraviolet and Visual
Echelle Spectrograph (UVES) on the European Southern
Observatory's (ESO) Very Large Telescope (VLT) facility
in 2002 (Program ID: 68.B-0115(A), PI: P.~Molaro),
and then again in 2007 to obtain UV coverage down to the
Lyman limit (Program ID: 078.A-0185(A), PI: M.~Pettini),
resulting in a total exposure time of 77\,500\,s. 
The details of this system's chemical composition
were reported by \citet{Pet08a} (see also \citealt{Ern06}),
and the analysis of the D\,{\textsc i} lines was presented 
by \citet{Pet08b}; with [O/H]\,$= -2.40$ this
is currently the most metal-poor DLA known with resolved \DI\ lines.
There is a small column density of ionized gas at velocities near that
of the DLA;  we have carefully modelled 
its contribution to the overall absorption
in the metal lines and the \HI\ Lyman
series (see Figure~\ref{fig:Q0913p072}).

Since Q0913$+$072 does not have an SDSS spectrum from which we
could perform a flux calibration, we obtained a set of VLT+FORS2
(Focal Reducer/low dispersion Spectrograph 2)
exposures from the VLT archive to flux calibrate the UVES spectrum.
(Program ID: 70.A-0425(C), PI: L.~Wisotzki). The FORS2 data
were reduced following standard procedures, using the
software routines from the ESO FORS2 data reduction
pipeline\footnote{We used version 3.9.6, obtained
from:\\ \texttt{http://www.eso.org/sci/software/pipelines/}}.
The UVES data and the corresponding model fits are shown in
Figure~\ref{fig:Q0913p072}.

\subsubsection{J1419$+$0829, $z_{\rm abs}=3.04973$}

The chemical properties of this DLA were discussed by \citet{Coo11},
while \citet{PetCoo12} reported the analysis of the deuterium lines. 
This system currently holds the record for the most precise
determination of the D/H ratio. The analysis technique 
developed in  \citet{PetCoo12} is the same as that employed here
(see Section~\ref{sec:analysis}), except for two aspects:
(1) in the present work all QSO spectra were analysed ``blind,''
i.e., without knowledge of the value of D/H until
the line fitting procedure was concluded; and
(2) we now allow for an arbitrary continuum fitting with
Legendre polynomials, rather than the combination
of power-law continuum and Gaussian
emission lines used by \citet{PetCoo12}.
To maintain consistency with the other DLAs investigated here,
we have reanalyzed this system with the new prescription for
the continuum definition, and the analysis was performed blind.
We found that the final value of D/H from this new analysis
deviated by just 0.005 dex from that reported by \citet{PetCoo12},
with an identical estimate for the error. This small shift is due to
our new approach for dealing with the systematic uncertainty
in the continuum level.

\subsubsection{J1558$-$0031, $z_{\rm abs}=2.70242$}

The data for J1558$-$0031  consist of a 11\,300\,s
Keck+HIRES spectrum obtained in 2006 (Program ID: U152Hb, PI: J.~X.~Prochaska),
first reported by \citet{OMe06}. 
With $\log N$(H\,{\textsc i})/cm$^{-2} = 20.67 \pm 0.05$,
this $z_{\rm abs}=2.70242$
DLA is the highest \HI\ column
density system currently known with at least one optically thin \DI\
absorption feature. There exists another metal-poor DLA along this
sightline, at $z_{\rm abs}=2.629756$, that also exhibits damped \Lya\ absorption
(log\,\NHI/cm$^{-2}$ = 19.726$\pm0.007$).
This second system contains a single absorption component, as evidenced
by several \SiII\ absorption lines 
(see bottom-right panel of Figure~\ref{fig:J1558m0031}).
Although \DI\ absorption is only seen 
in the higher redshift DLA, we include the \HI\ and
metal lines of both DLAs in the fitting procedure, 
since the damping wings of the two DLAs
overlap slightly. 
Portions of the HIRES spectrum, together with the best-fitting model, 
are reproduced in
Figure~\ref{fig:J1558m0031}.
Unlike the \citet{OMe06} analysis, which uses data acquired
with both Magellan+MIKE and Keck+HIRES, our analysis
exclusively uses the Keck+HIRES data described above.
\\

%%%%%%%%%%%%
% FIGURE 2 %
%%%%%%%%%%%%
\begin{figure*}
  \centering
  \includegraphics[angle=0, width=17.0cm]{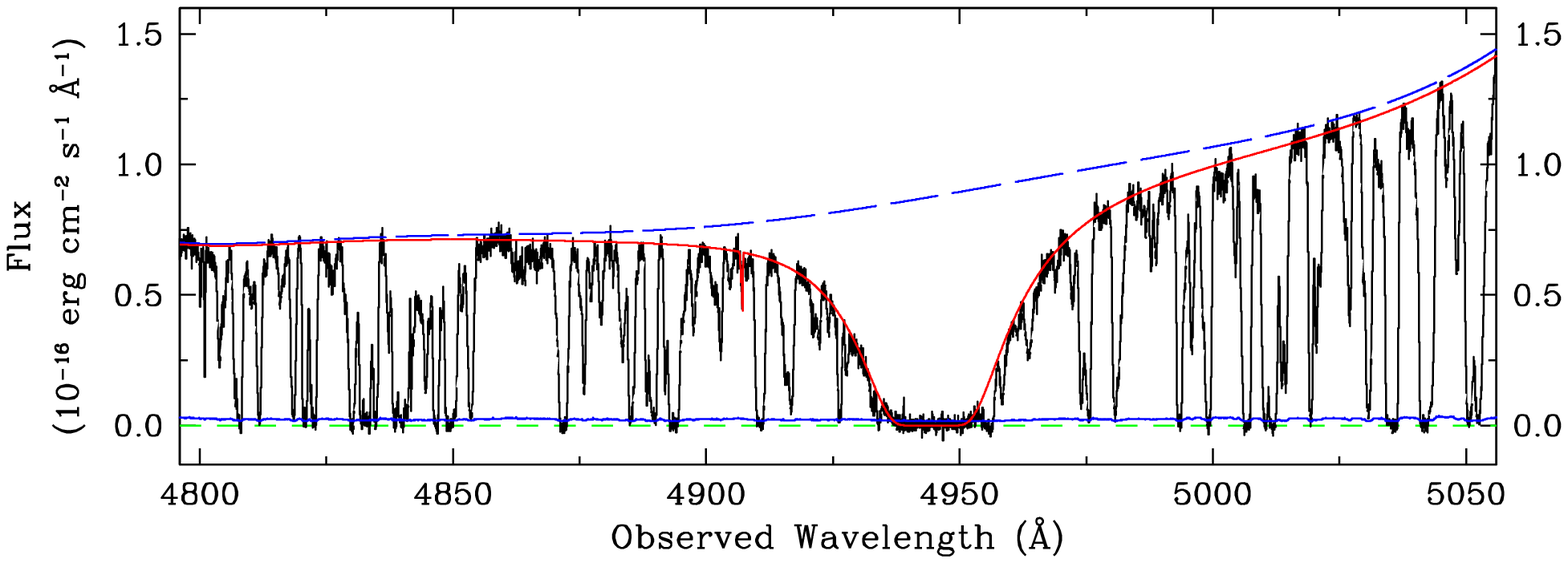}
  \includegraphics[angle=0, width=17.0cm]{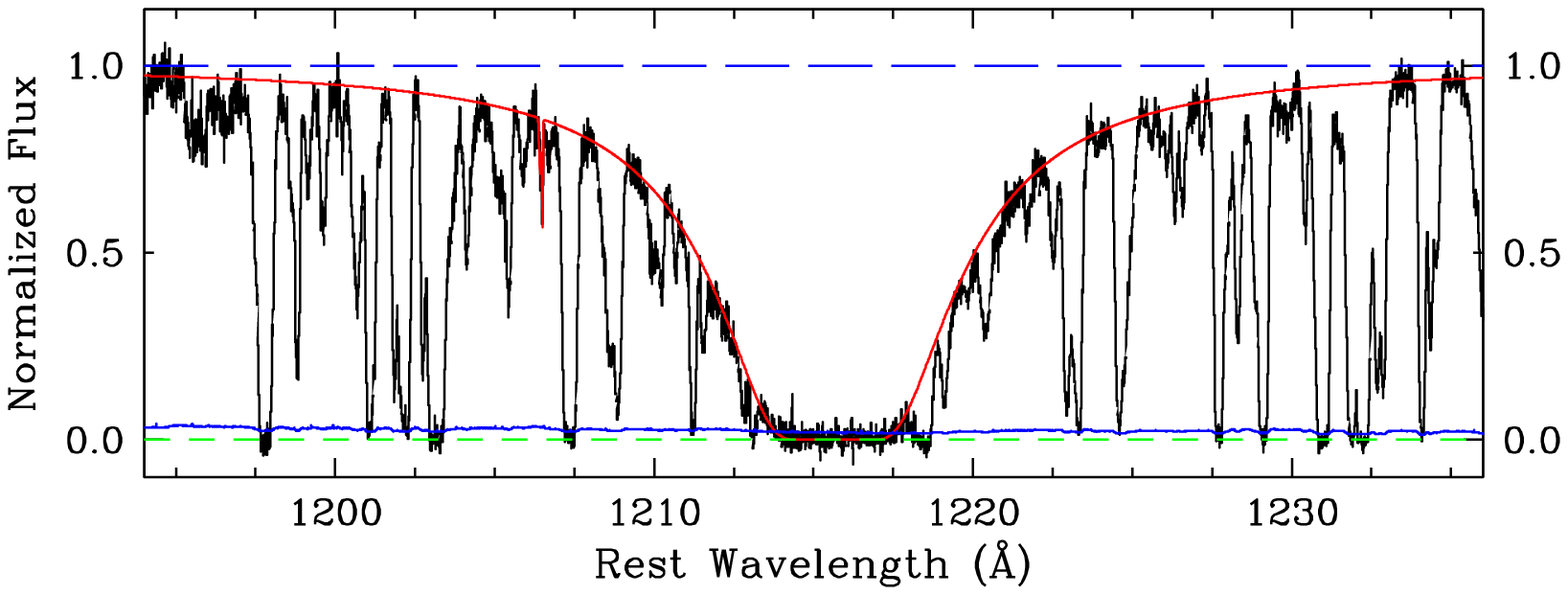}
  \caption{
The top panel displays a portion of the flux-calibrated HIRES 
spectrum near the damped
\Lya\ line at $z_{\rm abs}=3.06726$ toward
J1358$+$6522 (black histogram) together with the error spectrum
(continuous blue line). The dashed green line marks the best-fitting
zero level of the data and the dashed blue line shows the best-fitting
continuum level. The solid red line shows the overall best-fitting model
to the DLA. The bottom panel shows a zoom-in of the data and model; 
the weak absorption feature that we have modelled on the blue wing of
Ly$\alpha$ is \SiIII\,$\lambda 1206.5$ at the redshift of the DLA
($\sim4907$\,\AA\ in the observed frame).
 }
  \label{fig:lya}
\end{figure*}

\section{Profile Analysis}
\label{sec:profanalysis}

% Table of Component structure
\begin{table*}
\centering
\begin{minipage}[c]{0.75\textwidth}
    \caption{\textsc{Best-fitting Model Parameters for the DLA at $z_{\rm abs}=3.067259$ toward the QSO J1358$+$6522}}
    \begin{tabular}{@{}crrrcccc}
    \hline
    \hline
   \multicolumn{1}{c}{Comp.}
& \multicolumn{1}{c}{$z_{\rm abs}$} 
& \multicolumn{1}{c}{$T$} 
& \multicolumn{1}{c}{$b_{\rm turb}$} 
& \multicolumn{1}{c}{$\log N$\/(H\,{\sc i})}
& \multicolumn{1}{c}{$\log {\rm (D\,\textsc{i}/H\,\textsc{i})}$}
& \multicolumn{1}{c}{$\log N$\/(C\,{\sc ii})}
& \multicolumn{1}{c}{$\log N$\/(N\,{\sc i})}\\
    \multicolumn{1}{c}{}
& \multicolumn{1}{c}{}
& \multicolumn{1}{c}{(K)}
& \multicolumn{1}{c}{(km~s$^{-1}$)}
& \multicolumn{1}{c}{(cm$^{-2}$)}
& \multicolumn{1}{c}{}
& \multicolumn{1}{c}{(cm$^{-2}$)}
& \multicolumn{1}{c}{(cm$^{-2}$)}\\
  \hline

1a  & $3.0672594$               & $5\,700$                   & $2.4$          & $20.21$         & $-4.587^{\rm a}$         &  $14.32$                &  $12.86$                \\
    &  $\pm 0.0000005$       & $\pm 1\,000$           &$\pm 0.2$   & $\pm 0.06$    & $\pm 0.012$                &  $\pm  0.06$          &  $\pm  0.06$          \\
1b  & $3.0672591$               & $4\,000$                   & $8.8$          & $20.16$         & $-4.587^{\rm a}$         &  \ldots$^{\rm b}$   &  \ldots$^{\rm b}$   \\
    &  $\pm 0.0000018$       & $\pm 500$               &$\pm 0.5$   & $\pm 0.07$     & $\pm 0.012$                &                                 &                                 \\
2  & $3.06702$               & 13\,100                     & $8.5$          & $17.4$            & $-4.587^{\rm a}$         &  $12.30$                &  \ldots$^{\rm b}$   \\
    &  $\pm 0.00001$               &   $\pm1\,500$               &$\pm 1.4$      & $\pm 0.03$        & $\pm 0.012$                &  $\pm0.15$            &                  \\
  \hline
    \end{tabular}

    \smallskip\smallskip\smallskip

    \hspace{0.5cm}\begin{tabular}{@{}ccccccc}
    \hline
    \hline
   \multicolumn{1}{c}{Comp.}
& \multicolumn{1}{c}{$\log N$\/(O\,{\sc i})}
& \multicolumn{1}{c}{$\log N$\/(Al\,{\sc ii})}
& \multicolumn{1}{c}{$\log N$\/(Si\,{\sc ii})}
& \multicolumn{1}{c}{$\log N$\/(Si\,{\sc iii})}
& \multicolumn{1}{c}{$\log N$\/(S\,{\sc ii})}
& \multicolumn{1}{c}{$\log N$\/(Fe\,{\sc ii})}\\
    \multicolumn{1}{c}{}
& \multicolumn{1}{c}{(cm$^{-2}$)}
& \multicolumn{1}{c}{(cm$^{-2}$)}
& \multicolumn{1}{c}{(cm$^{-2}$)}
& \multicolumn{1}{c}{(cm$^{-2}$)}
& \multicolumn{1}{c}{(cm$^{-2}$)}
& \multicolumn{1}{c}{(cm$^{-2}$)}\\
  \hline

1a  & $14.85$                  & $11.93$                & $13.44$                &  $11.87$      &  13.00          & $13.09$ \\
    & $\pm 0.02$            & $\pm 0.03$           & $\pm 0.02$           &  $\pm0.04$  &  $\pm0.08$ & $\pm 0.02$ \\
1b  & \ldots$^{\rm b}$    & \ldots$^{\rm b}$   & \ldots$^{\rm b}$    &  \ldots$^{\rm b}$      & \ldots$^{\rm b}$  & \ldots$^{\rm b}$ \\
    &                                  &                                &                                &                       &   &  \\
2  & \ldots$^{\rm b}$     & \ldots$^{\rm b}$   & $11.53$                &  $11.80$      & \ldots$^{\rm b}$  & \ldots$^{\rm b}$ \\
    &                                  &                                & $\pm0.06$            &  $\pm0.06$  &   &   \\
  \hline
    \end{tabular}

    \smallskip
    
\hspace{0.5cm}$^{\rm a}${Forced to be the same for all components.}

\hspace{0.5cm}$^{\rm b}${Absorption is undetected for this ion in this component.}\\
    
    \label{tab:compstruct}
\end{minipage}
\end{table*}

The analysis we have adopted to measure the D/H ratio is
an improvement over previous studies, with two main
differences:
(1) the profile fitting and $\chi^2$ minimization process was
performed with new, purpose-built software routines; and (2) we
have tried to identify the dominant systematics, and have
attempted to account for these affects. Before we discuss the details
of our line profile analysis, we briefly describe the important aspects
of the new software we have written to measure D/H 
in metal-poor DLAs.
We will focus our discussion on the newly discovered system
with a precise value of the D/H abundance
(J1358$+$6522), but we stress that all five DLAs were analyzed in
an identical manner.

\subsection{Line Profile Fitting Software}

To account for the dominant systematics that could potentially affect
the measurement of D/H, we have developed a new suite of
software routines written in the \textsc{python} software environment.
Our Absorption LIne Software (\alis; described in more detail by
R.~Cooke 2014, in preparation) uses a modified version of the \textsc{mpfit}
package \citep{Mar09}. \textsc{mpfit} employs a Levenberg-Marquardt
least-squares minimization algorithm to derive the model parameters
that best fit the data (i.e. the parameters that minimize the difference
between the data and the model, weighted by the error in the data).
Unlike the software packages that have previously been used to
measure D/H, \alis\ has the advantage of being able to
fit an arbitrary emission profile for the quasar whilst simultaneously
fitting to the absorption from the DLA; any uncertainty in the continuum
is therefore automatically folded into the final uncertainty in the D/H ratio.
For all transitions, we use the atomic data compiled by \citet{Mor03}.

%%%%%%%%%%%%
% FIGURE 3 %
%%%%%%%%%%%%
\begin{figure*}
  \centering
  \includegraphics[angle=0, height=21.65cm]{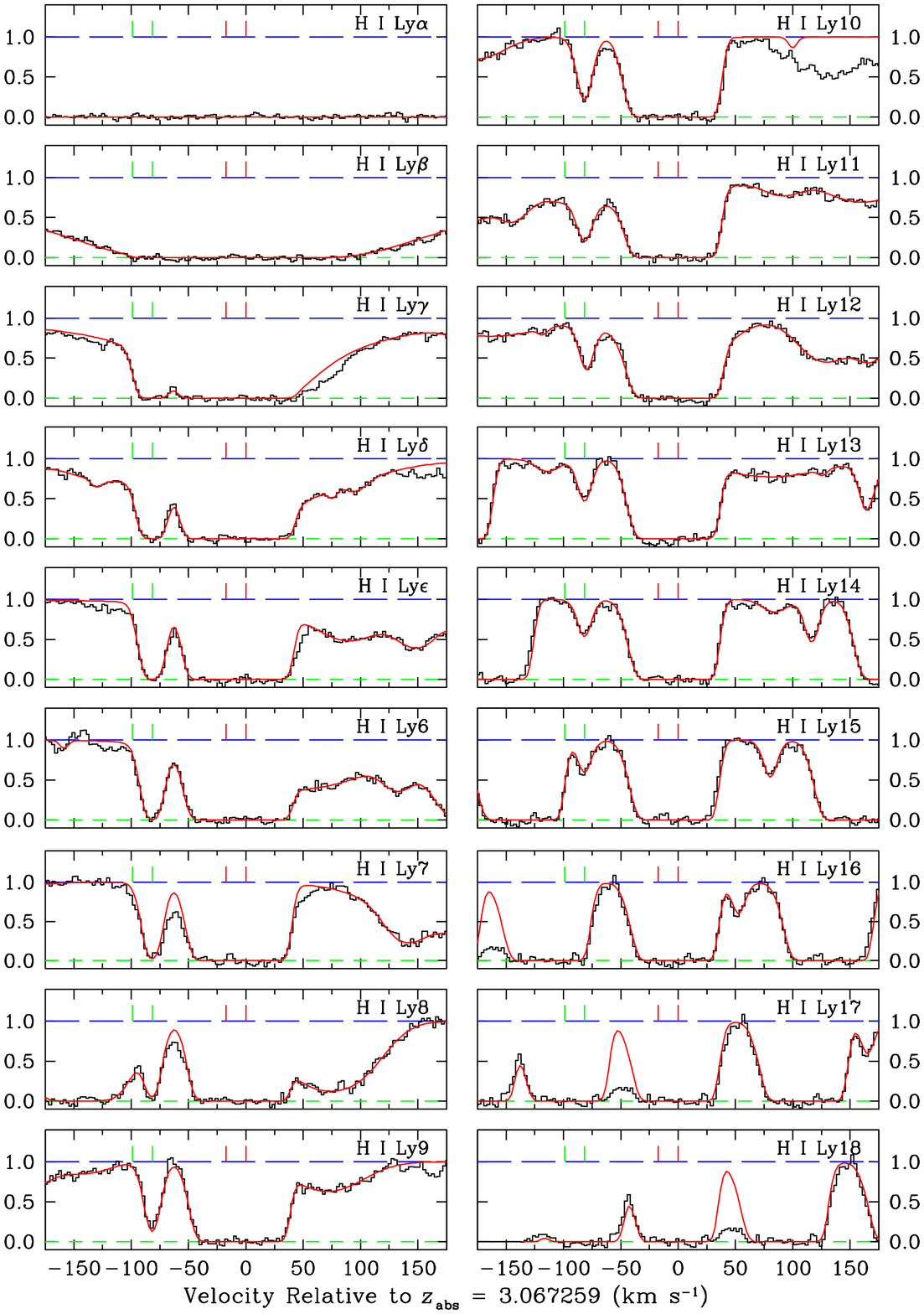}
  \caption{ 
A montage of the full Lyman series absorption in the DLA 
at $z_{\rm abs}=3.067259$
toward J1358$+$6522. 
The black histogram shows the data,
fully adjusted to the best-fitting continuum and zero levels, 
while the red continuous line is 
the model fit. 
The minimum $\chi^{2}$/dof for this fit is $6282.3/6401$.
Tick marks above the spectrum indicate
the location of the velocity components 
(red ticks for \HI, green ticks for \DI).
 }
  \label{fig:lyman}
\end{figure*}

\subsection{Analysis Technique for Measuring D/H}
\label{sec:analysis}

%%%%%%%%%%%%
% FIGURE 4 %
%%%%%%%%%%%%
\begin{figure*}
  \centering
  \includegraphics[angle=0, width=0.99\textwidth]{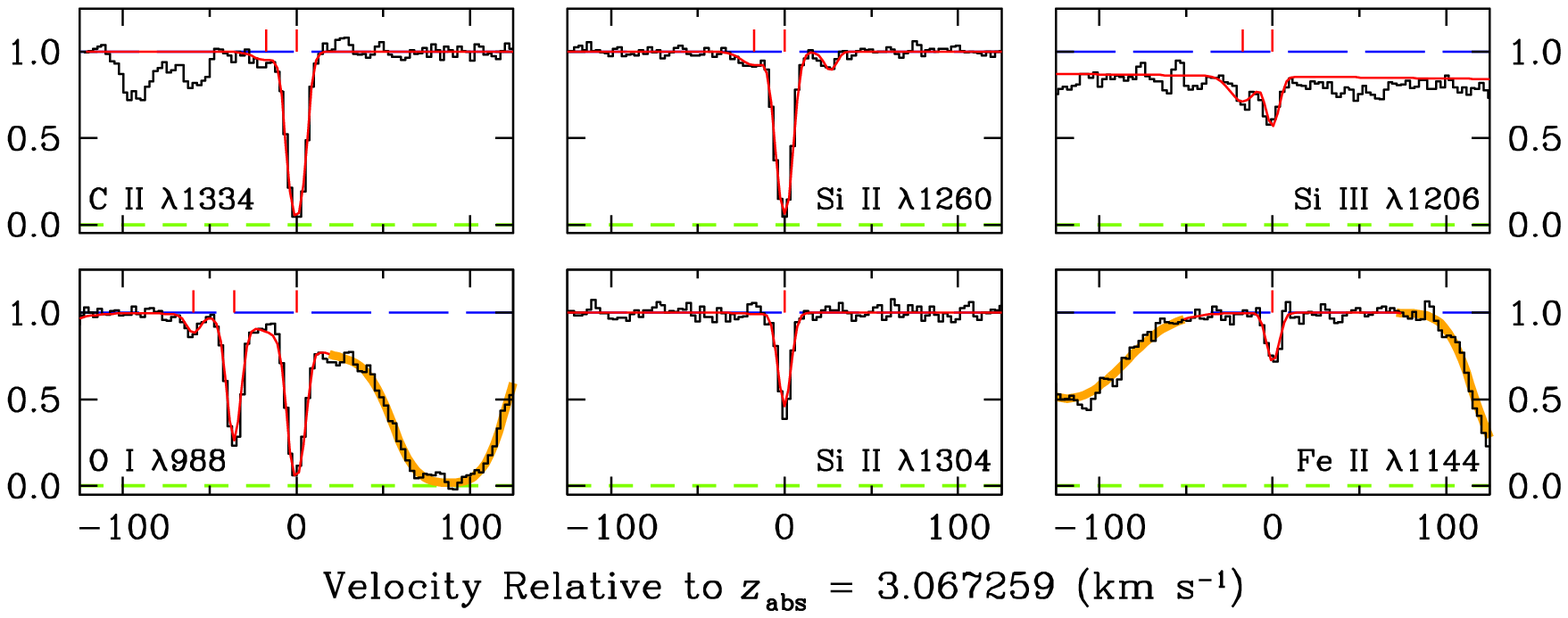}
  \caption{ 
A selection of metal absorption lines for the DLA at $z_{\rm abs}=3.067259$
toward the QSO J1358$+$6522 (black histogram, cf. \citealt{Coo12}). The
best-fitting model is shown as the solid red line, while fitted blends are shown
as the thick orange lines. In the top panels, the red tick marks above the spectrum
indicate the location of the two absorption components, while the bottom panels
only display the tick mark for the main absorption component (i.e. component 1a+1b).
The long-dashed blue and dashed green lines show the continuum and zero level
respectively. The absorption feature near $+25$ km s$^{-1}$
in the \SiII\,$\lambda1260$ panel is \FeII\,$\lambda1260.5$.
 }
  \label{fig:metals}
\end{figure*}

We first highlight one of the main strengths of our analysis which,
to our knowledge, has not been implemented in previous D/H studies. 
In order to remove human bias from the results, 
we have adopted a blind analysis strategy,
whereby the only information revealed during the profile fitting process is the
best-fitting $\chi^2$ value, and plots of the corresponding model fits
to the data; all of the parameter fitting results are entirely hidden from view.
Only after the analysis converged on the best-fitting model 
were the full results of the model fit uncovered, and 
these are the final values quoted here. No
further changes were made to the model fit
nor data reduction after the results were
revealed.

To allow for an efficient computation, a small wavelength interval
(typically $\pm300$ km s$^{-1}$) around each absorption line of interest
was extracted. For each of these intervals, we manually identified the
regions to be used in the fitting process by selecting pixels that we
attribute to either the quasar continuum or the DLA's absorption
lines. In some cases, it was also necessary to model unrelated
absorption lines in the fitting procedure if there was significant line-blending.

We have also attempted to account for the main systematics and limitations
that might affect the reported D/H value. The dominant systematics likely include
the instrumental broadening function, the cloud model, and the placement
of the quasar continuum and zero-level. We marginalize over the first of these
systematics
by assuming that the instrumental profile is a Gaussian\footnote{We confirmed that
emission lines from the ThAr wavelength calibration spectrum were well-fit by a
Gaussian profile.}, where the Gaussian FWHM
was allowed to be a free model parameter during the
$\chi^2$ minimization process.

Any uncertainty or bias in the choice of the cloud model is overcome by our
selection criteria listed in Section~\ref{sec:criteria}. 
Specifically, DLAs
have two advantages over other absorption line systems
for the determination of the D/H ratio, both stemming 
from their high \HI\ column densities:
(1) The \HI\ \Lya\ line exhibits well-defined Lorentzian damping
wings\footnote{\citet{Lee13} recently suggested that
the wings of the damped \Lya\ absorption line exhibit an
asymmetric deviation from a Lorenzian. Such a deviation
from a Lorentzian profile would be marginalized over during
the arbitrary continuum fitting process described below, and
is thus expected to have a negligible impact on our results.}
from which  the total \HI\ column
density can be deduced precisely and 
independently of the cloud model; and 
(2) many \DI\ transitions in the Lyman series are normally
detectable. With their widely varying 
oscillator strengths these multiple \DI\ lines can pin down 
the \DI\ column density very precisely;
the availability of unsaturated, high-order lines
ensures that the value of D/H thus deduced does
not depend on the detailed kinematic structure
of the gas (i.e. the cloud model).
Neither of these two advantages applies to Lyman limit
systems whose \HI\ column density can be three
orders of magnitude lower than that of DLAs.

Finally, if the background level is not accurately determined, 
a small offset can systematically
affect the derived D/H value. 
We have overcome this concern by allowing the zero level
to be fit during the minimization process
(assuming the offset to be the same at all wavelengths).

Arguably, the above systematic effects 
are sub-dominant compared to the choice of the
continuum level. One of the improvements 
over previous analyses is that we have not
specified \textit{a priori} the level of the quasar continuum.
Instead, we have marginalized
over the continuum level during the fitting process 
by fitting Legendre polynomials to the
flux-calibrated quasar continuum. 
The order of the polynomial was chosen to be sufficiently
large to allow flexibility in arbitrarily defining the continuum level, 
whilst ensuring that the continuum was not being over-fit to the noise
(examples are shown in the top panels of Figs.~\ref{fig:lya},
\ref{fig:HS0105p1619}, \ref{fig:Q0913p072}, and \ref{fig:J1558m0031}).

We emphasize that by including the above systematics into the fitting process,
the uncertainties in the continuum, zero level, and instrumental
broadening are incorporated
into the final uncertainty in the derived D/H value. 
To summarize, we have
marginalized over the continuum, zero level and instrumental
FWHM, whilst simultaneously fitting for all of the DLA's 
unblended metal lines, the DLA's
\Lya\ transition, and the entire \HI\ and \DI\ Lyman series. 
Our analysis therefore includes all of the available information 
that can help pin down the
D/H value,  and it accounts for the systematics that
are currently thought to contribute the most to the
overall error budget. 
We finally point out that every \HI\ component used
to fit the DLA was forced to have the same \DI\,/\,\HI\ ratio.
We have therefore fit directly to the \DI\,/\,\HI\ value, 
rather than deriving separately the column densities of \DI\ and \HI.
All absorption lines were modeled as Voigt profiles 
where the line broadening has both
thermal and turbulent contributions.
All of the DLAs in our survey were first modeled
with a single absorption component. If the metal lines
exhibited an asymmetric profile, an additional component
was included to improve the quality of the fit.

To confirm that the model parameters had
converged to their best-fitting values, we performed
a convergence test; once the difference in the
$\chi^{2}$ between successive iterations was $<0.01$,
the parameter values were stored and the $\chi^{2}$
minimization recommenced with a tolerance of $10^{-3}$.
Once a successive iteration reduced the $\chi^{2}$ by
$<10^{-3}$, the minimization was terminated and the
parameter values from the two convergence
criteria were compared. If all parameter values differed by
$<0.2\sigma$ (i.e. 20 per cent of the parameter error) then
the model fit has converged.

As a final step, we repeated the $\chi^{2}$ minimization process 20 times,
perturbing the starting parameters of each run by the covariance matrix. This
exercise ensures that our choice of starting parameters does not influence the
final result. We found that the choice of starting parameters has a negligible
contribution to the error on \DI\,/\,\HI\ (typically 0.002 dex), but can introduce
a small bias (again, typically 0.002 dex). We have accounted
for this small bias in all of the results quoted herein.

\subsection{Component Structure}
\label{sec:compstruc}

Most of the narrow, low-ionization metal lines of the DLA toward
J1358$+$6522 consist of a single component at
$z_{\rm abs}=3.067259$. 
A second weaker component,
blueshifted by $17.4$ km s$^{-1}$ ($z_{\rm abs} = 3.06702$)
contributes to 
\SiIII~$\lambda 1206.5$ and to the strongest \CII\ and \SiII\
lines. Evidently, this weaker absorption arises in nearby ionized gas. 

In fitting the absorption lines, we tied the redshift, turbulent Doppler parameter
and kinetic temperature of the gas to be the same for the metal,
\DI\ and \HI\ absorption lines. We allowed all of the cloud model
parameters to vary, whilst simultaneously fitting for the continuum
near every absorption line. Relevant parameters of the
best-fitting cloud model so determined are 
collected in Table~\ref{tab:compstruct}. 
Figures~\ref{fig:lya}, \ref{fig:lyman} and \ref{fig:metals}
compare the data and model fits for, respectively, the damped
\lya\ line, the full Lyman series, and selected metal lines.
[Since the metal lines analyzed here are the same as those
shown in Figure~1 of \citet{Coo12}, albeit now with
a higher S/N ratio, we only present a small selection of them in
Figure~\ref{fig:metals} to avoid repetition].
The best-fitting chi-squared value for
this fit is also provided for completeness\footnote{We
caution that the quoteed chi-squared value is likely underestimated
in our analysis because: (1) there is some degree of correlation
between neighbouring pixels that is not accounted for in the
error spectrum, and (2) the continuum regions selected tend
to have lower fluctuations about the mean than average.}.

%%%%%%%%%%%%
% FIGURE 5 %
%%%%%%%%%%%%
\begin{figure*}
  \centering
  {\hspace{-0.25cm} \includegraphics[angle=0,width=8.5cm]{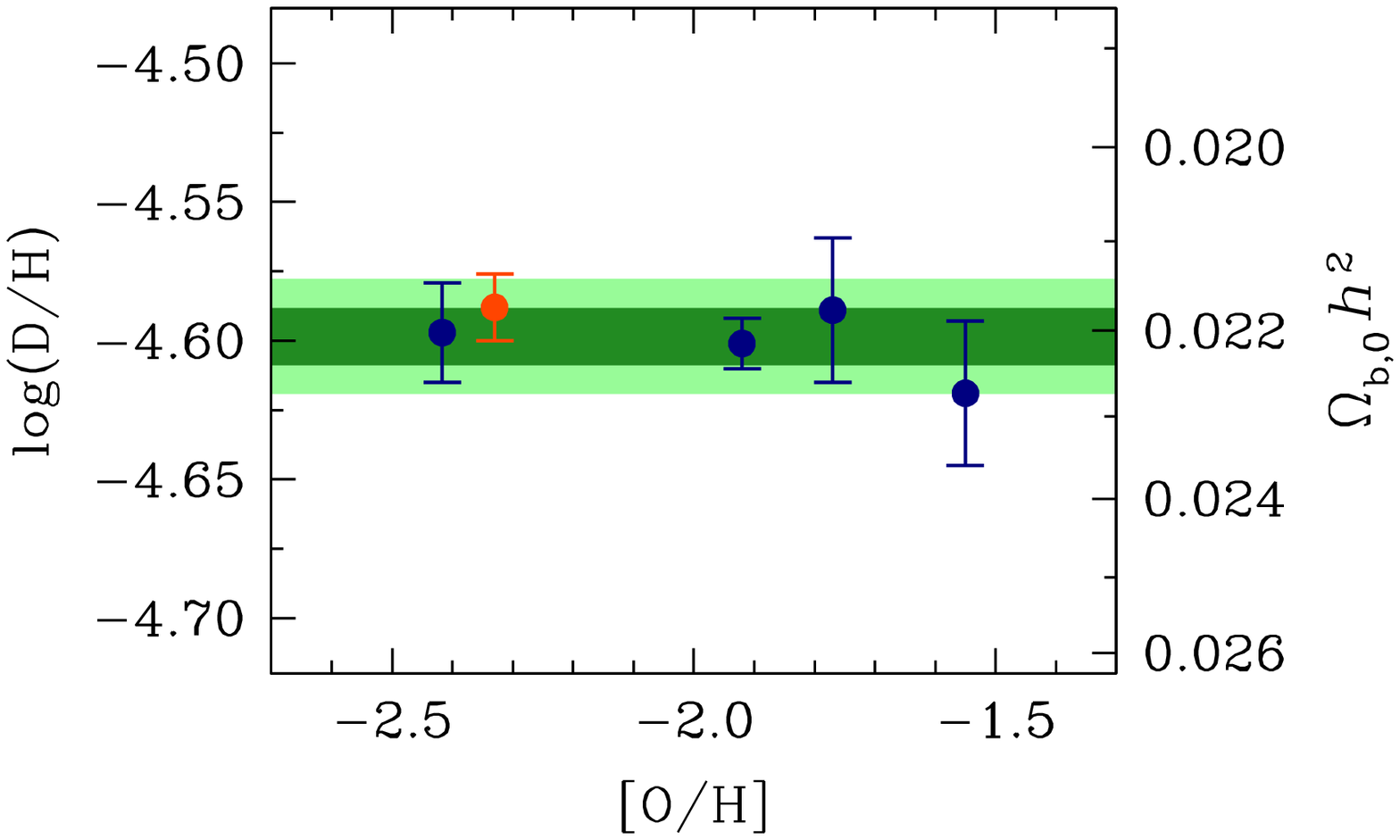}}
 {\hspace{0.25cm} \includegraphics[angle=0,width=8.5cm]{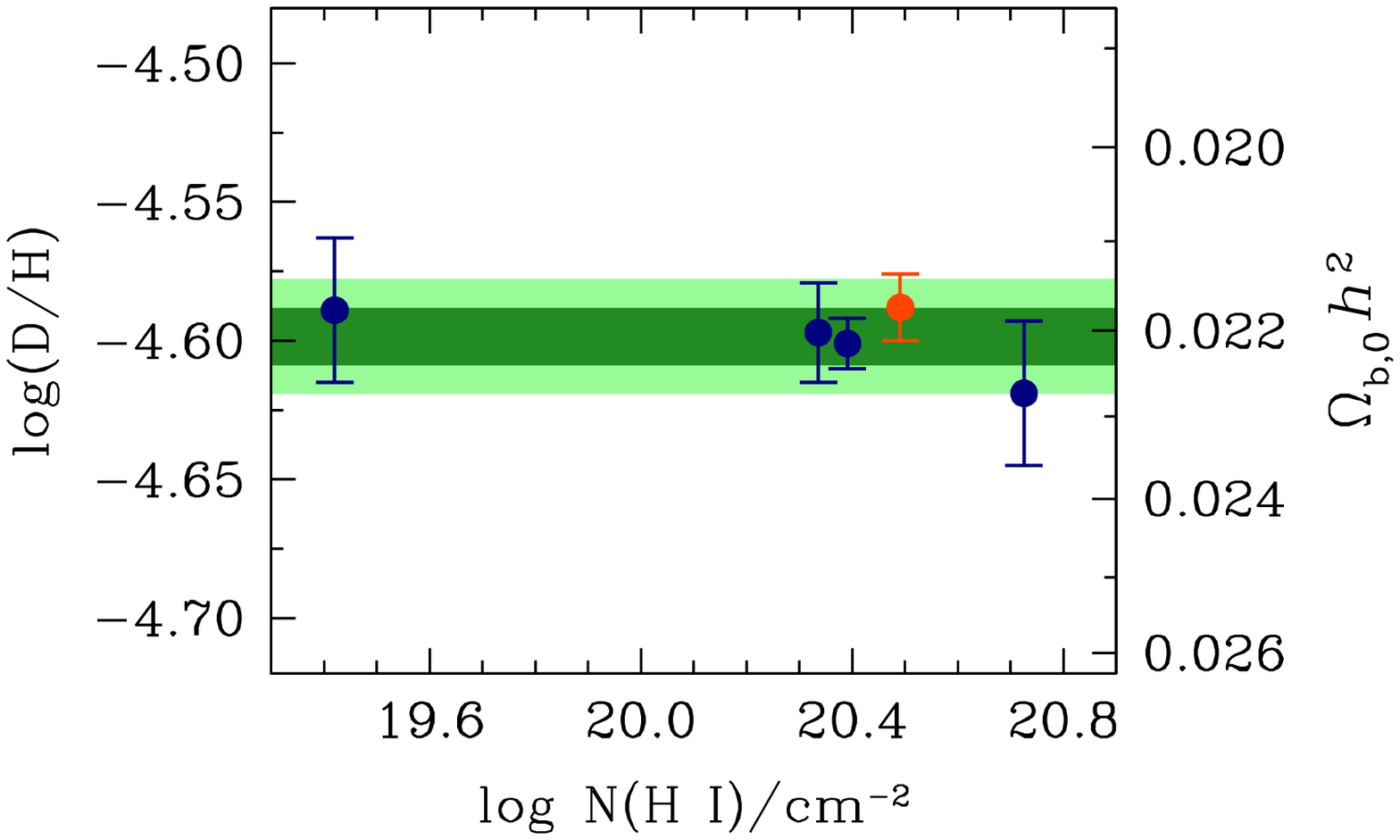}}
  \caption{ 
Values of D/H for the Precision Sample of DLA measurements 
analyzed in this paper. The orange
point represents the new case reported here (J1358+6522). The left and right panels
show respectively the D/H measures as a function of the DLA oxygen abundance and
\HI\ column density. The dark and light green bands are the 1$\sigma$ and
2$\sigma$ determinations of \Obary\hsq\  from the analysis
of the CMB temperature fluctuations recorded by the \textit{Planck}
satellite \citep{Efs13} assuming the standard model of physics.
The conversion from D/H to \Obary\hsq\ is given by eqs.~\ref{eqn:dhptoetad}
and \ref{eqn:etad}.
 }
  \label{fig:dhsample}
\end{figure*}

Returning to Table~\ref{tab:compstruct}, it can be seen
that we found it necessary to separate the main absorption
into two separate components,
labeled 1a and 1b in the Table.
A statistically acceptable fit\footnote{The addition of an extra absorption
component (i.e. three components as opposed to two, and four additional
free parameters) reduces the minimum
chi-squared value by $\Delta\chi^{2}_{\rm min}\simeq660$, which is
\textit{highly} significant (see e.g. \citealt{LamMarBow76}).} to the metal, \DI\ and \HI\
lines could not be achieved with a single absorbing cloud in which
the turbulent broadening is the same for all species and
the thermal broadening is proportional to the square root
of the ion mass (i.e. $b_{\rm th}^2 = 2KT/m$, where $K$
is the Boltzmann constant). 
The main absorption component of this DLA appears
to consist of two `clouds' with very similar redshifts,
temperatures and \HI\ column densities, but with 
significantly different turbulence parameters
(see Table ~\ref{tab:compstruct}).
The turbulent broadening for component
1a is bounded by the metal-lines, whereas the thermal
broadening is bounded by the relatively narrow \HI\
line profiles. This combination of turbulent and thermal
broadening is unable to reproduce the observed widths
of the strongest \DI\ lines, which require an additional
component with a larger contribution of turbulent broadening.
Surprisingly, metal absorption is only seen in the low-turbulence
cloud (component 1a); component 1b has less than 1/100
of the metallicity of component 1a.
We propose three possible interpretations for this unusual finding:
(1) there exists an essentially pristine cloud of gas near the
very metal-poor DLA; (2) the metals see a different potential
to \DI\ and \HI\ (i.e. the metals have not been entirely mixed into
the \HI); (3) the assumption of a constant kinetic temperature for
all elements through the entire sightline is incorrect.

While any of these explanations is interesting in its own right,
we stress here
that \textit{the choice of the cloud model does not affect the derived
value of D/H} in this system, for the reasons discussed in
Section~\ref{sec:analysis}. Provided that there exists a series of
optically thin \DI\ lines, and \HI\ \Lya\ exhibits
damping wings
(as indeed is the case for the 
DLA towards J1358+6522---see Figures~\ref{fig:lya}
and \ref{fig:lyman}), the derived total \DI\ and \HI\ column densities will
be independent of the cloud model. We therefore only require
that the model provides a good fit to the data.

The final, best-fitting value of the deuterium abundance for the DLA
towards J1358$+$6522 is (expressed as
a logarithmic and linear quantity):
\begin{equation}
\log\,(\textrm{D}\,\textsc{i}/\textrm{H}\,\textsc{i}) = -4.588 \pm 0.012
\end{equation}
\begin{equation}
10^5\,\textrm{D}\,\textsc{i}/\textrm{H}\,\textsc{i} = 2.58 \pm 0.07
\end{equation}
where the quoted error term includes the random (observational)
error in addition to the systematic uncertainties that were
marginalized over during the fitting process.

\section{The Primordial Abundance of Deuterium from the Precision Sample}
\label{sec:cosmology}

\begin{table*}
\begin{minipage}{0.9\textwidth}
    \caption{\textsc{The Precision Sample of D/H Measurements in QSO Absorption Line Systems}}
    \begin{tabular}{lccc|cc|cc|l}
    \hline
    \hline
\multicolumn{4}{c}{}
& \multicolumn{2}{c}{Literature}
& \multicolumn{2}{c}{This work}
& \multicolumn{1}{c}{}\\
   \multicolumn{1}{c}{QSO}
& \multicolumn{1}{c}{$z_{\rm em}$} 
& \multicolumn{1}{c}{$z_{\rm abs}$} 
& \multicolumn{1}{c}{[O/H]$^{\rm a}$}
& \multicolumn{1}{c}{$\log N$\/(H\,{\sc i})}
& \multicolumn{1}{c}{$\log {\rm (D/H)}$}
& \multicolumn{1}{c}{$\log N$\/(H\,{\sc i})}
& \multicolumn{1}{c}{$\log {\rm (D/H)}$}
& \multicolumn{1}{c}{Ref.$^{\rm b}$}\\
    \multicolumn{1}{c}{}
& \multicolumn{1}{c}{}
& \multicolumn{1}{c}{}
& \multicolumn{1}{c}{}
& \multicolumn{1}{c}{(cm$^{-2}$)}
& \multicolumn{1}{c}{}
& \multicolumn{1}{c}{(cm$^{-2}$)}
& \multicolumn{1}{c}{}
& \multicolumn{1}{c}{}\\    
  \hline
HS\,0105+1619                        & 2.652     &  2.53651   & $-1.77$  &   $19.42 \pm 0.01$        &  $-4.60 \pm 0.04$      &   $19.426 \pm 0.006$        &  $-4.589 \pm 0.026$  &  1, 2 \\
Q0913+072                               & 2.785     &  2.61829   & $-2.40$  &   $20.34 \pm 0.04$        &  $-4.56 \pm 0.04$      &   $20.312 \pm 0.008$        &  $-4.597 \pm 0.018$  &  1, 3, 4 \\
SDSS~J1358$+$6522            & 3.173     &  3.06726   & $-2.33$  &                  $\ldots$                    &                $\ldots$                   &   $20.495 \pm 0.008$        &  $-4.588 \pm 0.012$  &  1 \\
SDSS~J1419$+$0829            & 3.030     &  3.04973   & $-1.92$  &   $20.391 \pm 0.008$   &  $-4.596 \pm 0.009$  &   $20.392 \pm 0.003$   &  $-4.601 \pm 0.009$  &  1, 5, 6 \\
SDSS~J1558$-$0031             & 2.823     &  2.70242   & $-1.55$  &   $20.67 \pm 0.05$        & $-4.48 \pm 0.06$       &   $20.75 \pm 0.03$      &  $-4.619 \pm 0.026$  & 1, 7  \\
   \hline
    \end{tabular}
    \smallskip

$^{\rm a}${We adopt the solar value $\log ({\rm O/H})_{\odot} + 12 = 8.69$ \citep{Asp09}.}\\
$^{\rm b}${References -- (1)~This work,
(2)~\citet{OMe01}, 
(3)~\citet{Pet08a},
(4)~\citet{Pet08b},
(5)~\citet{PetCoo12},
(6)~\citet{Coo11},
(7)~\citet{OMe06}.
}\\
    \label{tab:precision}
\end{minipage}
\end{table*}

Including the new D\,/\,H case reported herein, there are just
five metal-poor absorption line systems in which a \textit{precise}
measure of the primordial abundance of deuterium can be obtained,
based on the criteria outlined in Section~\ref{sec:litsyst}.
Relevant details of these
systems are collected in Table~\ref{tab:precision}.
The four DLAs from the literature were all reanalyzed in
an identical manner to that described above (Section 3)
for J1358$+$6522. In particular, we adopted the same blind
analysis strategy and marginalized over the
important systematic uncertainties. We refer
to this sample of five high quality measurements
as the \textit{Precision Sample}.

In Table~\ref{tab:precision}, we provide a measure of
the \textit{total} \HI\ column density, along with the associated error.
Many of our systems contain more than one component in
\HI\ and the column density estimates for these multiple
components are correlated with one another.
To calculate the error on the total \HI\ column density,
we have drawn 10\,000 realizations of the component column
densities from the covariance matrix. We then calculated the total
column density for each realization; in Table~\ref{tab:precision},
we provide the mean and $1\sigma$ error derived from
this Monte Carlo analysis.

We consider the five measures of \DI\,/\,\HI\ in these DLAs
as five independent determinations of the primordial
abundance of deuterium, \dhp, for the following reasons:
(1) we are not aware of any physical mechanism that would
alter the ionization balance of D compared to H. Thus, to our
knowledge, \DI\,/\,\HI\,$\equiv$\,D\,/\,H;
(2) the degree of astration of D (i.e. its destruction
when gas is turned into stars) is expected to be
negligible at the low metallicities ([O/H]\,$< -1.5$)
of the DLAs considered here (e.g., see 
Figure 2 of Romano et al. 2006);
thus (D\,/\,H)$_{\rm DLA}$ = (D\,/\,H)$_{\rm p}$.
(3) the lack of dust in metal-poor DLAs 
makes it extremely unlikely that 
selective depletion of D onto grains
occurs in the cases considered here
[it has been proposed that such a mechanism
may be responsible for the local variations in (D\,/\,H)$_{\rm ISM}$---see
Linsky et al. 2006];
(4) the five DLAs sample entirely independent
sites in the distant Universe.

As can be seen from Table~\ref{tab:precision}
and Figure~\ref{fig:dhsample}, the five measures of D/H
in the Precision Sample are in very good mutual agreement
and the dispersion of the measurements is consistent with
the errors estimated with our improved analysis. 
A $\chi^{2}$ test indeed confirms that the five measurements
are consistent within $2\sigma$ of being drawn from a single
value of D/H.
We can therefore combine the five independent determinations
of (D\,/\,H)$_{\rm DLA}$ to deduce the weighted mean value
of the primordial abundance of deuterium:
\begin{equation}
{ \log\,{\rm (D/H)}_{\rm p}}=-4.597\pm0.006
\label{eqn:primdh_log}
\end{equation}
\begin{equation}
\label{eqn:primdh}
10^5\,{\rm (D/H)_{\rm p}}=2.53\pm0.04
\end{equation}

This value of \dhp\ is not markedly different from 
other recent estimates 
\citep{Pet08b, FumOmePro11, PetCoo12},
but its precision is significantly better than 
achieved in earlier papers that considered a 
more heterogeneous set of 
${\rm (D/H)_{DLA}}$ determinations. 
For completeness, we have recalculated the weighted mean for
all the known D/H measurements listed in Table~2 of
\citet{PetCoo12}, after updating the D/H values of the systems we
have reanalyzed here. The resulting weighted mean value of the
primordial deuterium abundance is \dhp\,$ =-4.596\pm0.006$.
This compares very well with
the value derived from the Precision Sample (eqs.~\ref{eqn:primdh_log}
and \ref{eqn:primdh}).
Perhaps this is not surprising, since
the literature systems that
did not meet our selection criteria (see Section~\ref{sec:criteria})
have larger uncertainties, and thus their contribution to the weighted mean
value of D\,/\,H is relatively low.

\subsection{The Cosmic Density of Baryons}
\label{sec:omega_b}

Using the most up-to-date calculations of the network of
nuclear reactions involved in BBN,
the primordial abundance of deuterium is related to the 
cosmic density of baryons (in units of the critical density),
\Obary,  via the following relations
(\citealt{Ste12}; G.~Steigman 2013, private communication):
\begin{equation}
\label{eqn:dhptoetad}
{\rm (D\,/\,H)}_{\rm p} = 2.55\times10^{-5}\,(6/\eta_{\rm D})^{1.6}\times(1\pm0.03)
\end{equation}
\begin{equation}
\label{eqn:etad}
\eta_{\rm D} = \eta_{\rm 10} - 6(S-1) + 5\xi/4
\end{equation}
where $\eta_{10}=273.9\,\Omega_{\rm b,0}\,h^{2}$, 
$S = [1 + 7(N_{\rm eff}-3.046)/43]^{1/2}$ is the expansion factor
and $\xi$ is the neutrino degeneracy parameter (related to the
lepton asymmetry by Equation 14 from \citealt{Ste12}). The rightmost term in
eq.~\ref{eqn:dhptoetad} represents the current 3\%
uncertainty in the conversion of \dhp\ to $\eta_{\rm D}$
due to the uncertainties in the relevant nuclear reactions rates 
(see Section~\ref{sec:limitation}).
For the standard model, 
\neff$\,\simeq3.046$ and $\xi=0$.
In this case, the Precision Sample of D/H measurements
implies a cosmic density of baryons:
\begin{equation}
\label{eq:dhobary}
100\,\Omega_{\rm b,0}\,h^2 ({\rm BBN})
= 2.202\pm0.020~~(\textrm{random})~\pm0.041~~(\textrm{systematic})
\end{equation}
where we have decoupled the error terms from our measurement 
(i.e. the random error term)
and the systematic uncertainty in converting the D abundance into
the baryon density parameter. 

As can be seen from Figure~\ref{fig:dhsample},
this value of \Obary\hsq\ is in excellent agreement
with that derived from the analysis of the CMB temperature
fluctuations measured by the \textit{Planck} satellite \citep{Efs13}:
\begin{equation}
\label{eq:dhobary_CMB}
100 \, \Omega_{\rm b,0} \, h^2 {\rm (CMB)} =2.205 \pm 0.028 .
\end{equation}

\subsection{The Current Limitation}
\label{sec:limitation}

In the era of high-precision cosmology, we feel that it is important to
highlight the main limitations affecting the use
of \dhp\ in the estimation of cosmological parameters. 
As can be seen from eq.~\ref{eq:dhobary}, 
the main source of error is in the conversion of
\dhp\ to the baryon density parameter ($\eta_{\rm D}$,
and hence \Obary\hsq). 
In large part, this systematic uncertainty is due  to 
the relative paucity of experimental measures for several nuclear
cross-sections that are important in the network of 
BBN reactions, particularly deuteron--deuteron reactions
and the \dHe\ reaction rate at the relevant energies
\citep{Fio98, NolBur00, Cyb04, Ser04}.
Since these studies, estimates for
the deuteron--deuteron reaction cross-sections \citep{Leo06} have
improved and  their contribution to the error
budget has been reduced.
The main lingering concern involves the reaction rate \dHe,
for which only a single reliable dataset of the {\it S}-factor is
currently available in the relevant energy range
\citep{Ma97}.\footnote{To avoid possible confusion
caused by the unfortunate
nomenclature, we point out that the 
{\it S}-factor in question is directly related to the reaction
cross-section as a function of energy, $\sigma(E)$.
It has nothing to do with the expansion factor
$S$ in eq.~\ref{eqn:etad}.}
The concern for \dHe\ is made
worse by the fact that the theoretical and experimental
values of the {\it S}-factor do not agree.
This paucity of data, in addition to the difficulties of obtaining
an accurate and precise measure of \dHe\ at BBN energies, led
\citet{NolHol11} to adopt a theoretical curve for the {\it S}-factor
between 50--500\,keV. This has resulted in the improved
conversion given in eq.~\ref{eqn:dhptoetad}.

It is worth pointing out here that a further reduction
by a modest factor of two in the uncertainty of the conversion from
\dhp\ to $\eta_{\rm D}$ would make the precision of 
\Obary\hsq(BBN) from the DLA Precision Sample
comparable to that of  \Obary\hsq(CMB)  
achieved by the \textit{Planck} mission 
(cf. eqs.~\ref{eq:dhobary} and \ref{eq:dhobary_CMB}).

%%%%%%%%%%%%
% FIGURE 6 %
%%%%%%%%%%%%
\begin{figure*}
  \centering
  {\hspace{-0.25cm} \includegraphics[angle=0,width=8.8cm]{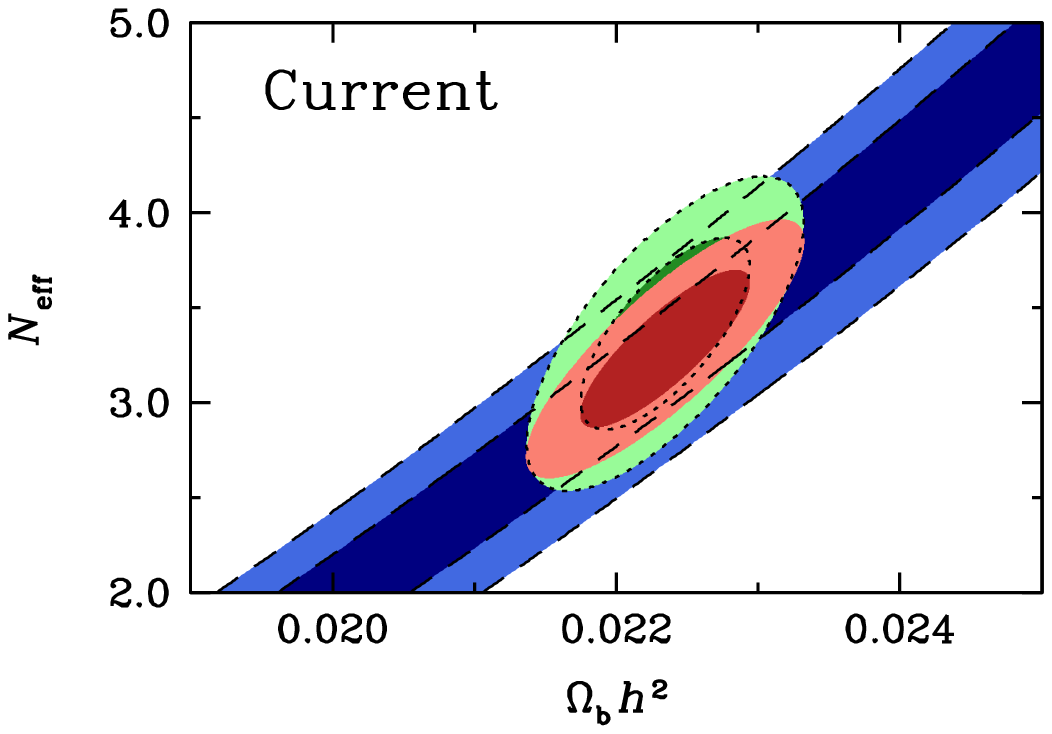}}
 {\hspace{0.25cm} \includegraphics[angle=0,width=8.8cm]{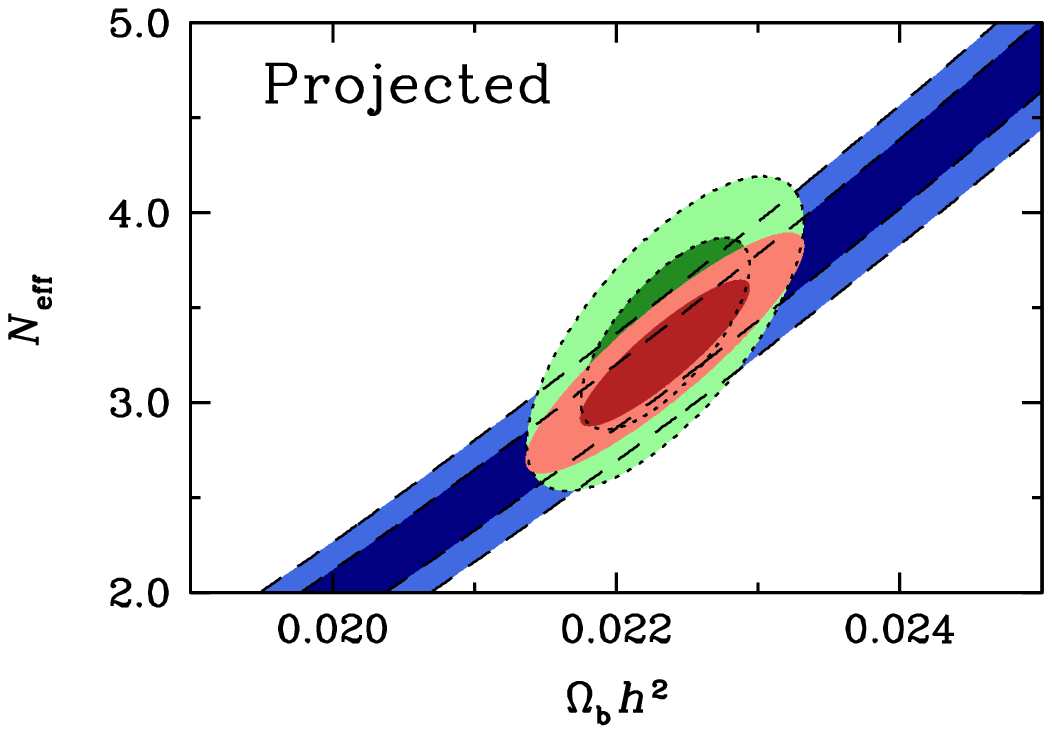}}
  \caption{ 
The $1\sigma$ and $2\sigma$ confidence contours (dark and light shades respectively)
for \neff\ and \Obary\hsq\ derived from the primordial 
deuterium abundance (blue), the
CMB (green), and the combined confidence contours (red). 
The left panel illustrates the current situation, 
while the right panel shows the effect of reducing the
uncertainty in the conversion from \dhp\ to \Obary\hsq\
by a factor of two (see discussion in Section~\ref{sec:limitation}).
Dashed and dotted lines indicate the
hidden contour lines for BBN and CMB bounds respectively.
 }
  \label{fig:neff}
\end{figure*}

\section{Beyond the Standard Model}
\label{sec:newphysics}

In this section we combine the results presented here
with those of the CMB analysis by the 
\textit{Planck} Collaboration to place bounds 
on some parameters denoting new physics 
beyond the standard model of cosmology and
particle physics. 
In particular, we use the Markov-Chain
Monte Carlo chains from the combined analysis of
the \textit{Planck} temperature fluctuations \citep{Efs13},
the low-multipole \textit{WMAP} polarization data
($l<23$; \citealt{Ben12}), and high-resolution CMB
data (see \citealt{Efs13} for an appropriate list of
references). Throughout, we refer to this combined
dataset as \textit{Planck}+WP+highL, for consistency
with the work by the \citet{Efs13}. In what follows, we
have assumed that \neff\ and the
baryon-to-photon ratio remained unchanged from the
epoch of BBN to the epoch of recombination.

\subsection{Combined CMB+BBN Measure of $N_{\rm eff}$ and \Obary\hsq}
\label{sec:neffobary}
It has long been known that the mass fraction of $^4$He, \yp, 
is potentially a very sensitive
probe of additional light degrees of freedom \citep{SteSchGun77}.
Unfortunately, systematic uncertainties in the determination
of \yp\ have limited its use as an effective probe
of \neff\ (see Figure~8 of \citealt{Ste12}).
However, \citet{NolHol11} (see also \citealt{Cyb04}) have
recently highlighted the potential of using precise measurements
of the primordial deuterium abundance in conjunction
with observations of the CMB to place a strong,
independent bound on \neff. 

In the left panel of Figure~\ref{fig:neff}, we show the 
current $1\sigma$ and $2\sigma$
confidence contours for \neff\ and \Obary\hsq\ derived from the
\textit{Planck}+WP+highL CMB analysis\footnote{We used the base
cosmology set with \neff\ added as a free parameter
(see Section 6.4.4 of \citealt{Efs13}).} (green), and from the
BBN-derived \dhp\ reported here (blue). 
The combined confidence bounds are
overlaid as red contours. 
In what follows, it is instructive to recall that
the CMB-only bounds are \citep{Efs13}:
\begin{equation}
100\,\Omega_{\rm b,0}\,h^2 = 2.23\pm0.04
\label{eq:dhobary_CMB_ns}
\end{equation}
\begin{equation}
N_{\rm eff} =  3.36\pm 0.34 \, .
\end{equation}
(Note that solving simultaneously for \Obary\hsq\
and \neff\  leads to a slightly different best-fitting
value of \Obary\hsq\ than that obtained
for the standard model; cf. eqs.~\ref{eq:dhobary_CMB} and 
\ref{eq:dhobary_CMB_ns}).
For comparison, from the joint BBN+CMB analysis
we deduce:
\begin{equation}
100\,\Omega_{\rm b,0}\,h^2 = 2.23\pm0.04\\
\label{eq:dhobary_ns}
\end{equation}
\begin{equation}
N_{\rm eff} =  3.28 \pm 0.28 \, .
\end{equation}

Thus, combining \dhp\ with the CMB does not significantly
change the uncertainty in \Obary\hsq, but does reduce the error
on \neff\ by $\sim20$ per cent. The uncertainty
on \neff\ could be reduced further by an improvement
in the cross-section of the \dHe\
(see right panel of Figure~\ref{fig:neff},
and Section~\ref{sec:limitation}).
Based on the current bound on \neff\ from CMB+\dhp, we can nevertheless
rule out the existence of an additional (sterile) neutrino
(i.e. \neff\, = 4.046) at $99.3\%$
confidence (i.e. $\sim 2.7\sigma$), 
provided that \neff\ and $\eta_{10}$
remained unchanged between
BBN and recombination. 
However, as noted recently by \citet{Ste13}, if the CMB photons
are heated after the neutrinos have decoupled [for example, by
a weakly interacting massive particle that annihilates
to photons], \neff\ will be less than $3.046$ for three standard model
neutrinos; a sterile neutrino can in principle exist even when \neff~$<4.046$.

Looking to the future, \yp\ has contours that are almost orthogonal
to those of the CMB and \dhp\ (see e.g. \citealt{Ste07}). Thus,
measures of \yp\ that are not limited by systematic uncertainties
could potentially provide a very strong bound,
when combined with \dhp, on the number of equivalent neutrinos
during the epoch of BBN, independently of CMB observations.

\begin{figure}
  \centering
  \includegraphics[angle=0,width=8.8cm]{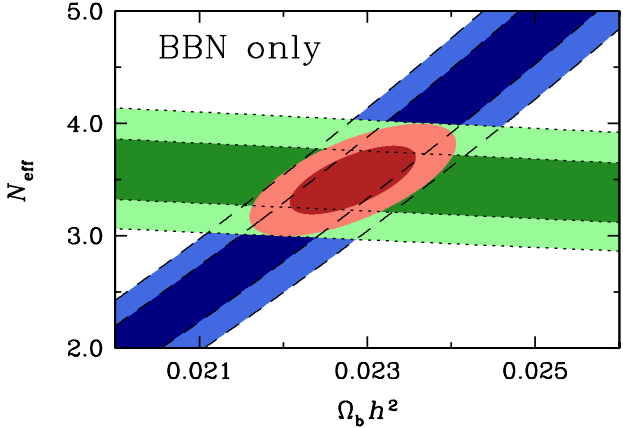}
  \caption{ 
The $1\sigma$ and $2\sigma$ confidence contours (dark and light shades respectively)
for \neff\ and \Obary\hsq\ derived from the primordial 
deuterium abundance (blue), the
primordial He mass fraction (green), and the combined confidence contours (red). 
Dashed and dotted lines indicate the
hidden contour lines for \dhp\ and \yp\ bounds respectively.
 }
  \label{fig:ypdhcont}
\end{figure}

Following improvements in the He\,{\textsc i} emissivity
calculations (Porter et al. 2012, 2013), there have been
two recent reassessments  of $Y_{\rm P}$, by
\citet{IzoStaGus13} and
by \citet{Ave13}  respectively.
\citet{IzoStaGus13} find $Y_{\rm P} = 0.253 \pm 0.003$;
this value includes a small correction of $-0.001$ to reflect
the fact that their \textit{Cloudy} modelling overpredicts
$Y_{\rm P}$ by this amount at the lowest metallicities --- see
their Figure~7(a). \citet{Ave13} find $Y_{\rm P} = 0.254 \pm 0.004$
from the average of all the low metallicity \HII\
regions in their sample.
Thus, the values deduced by these two teams
are in good mutual agreement; in the analysis that
follows we adopt $Y_{\rm P} = 0.253 \pm 0.003$
from \citet{IzoStaGus13}. Using the following conversion
relation for \yp\ (Steigman 2012; G.~Steigman 2013, private communication):
\begin{equation}
\label{eqn:yptoetad}
{Y}_{\rm P} = 0.2469 \pm 0.0006 + 0.0016\,(\eta_{\rm He} - 6)
\end{equation}
\begin{equation}
\label{eqn:etahe}
\eta_{\rm He} = \eta_{\rm 10} + 100(S-1) - 575\xi/4
\end{equation}
combined with the \citet{IzoStaGus13} measure of \yp,
we derive the following BBN-only bound on the 
baryon density and the effective number of neutrino
species:
\begin{equation}
100\,\Omega_{\rm b,0}\,h^2 = 2.28\pm0.05\\
\label{eq:dhobary_ns}
\end{equation}
\begin{equation}
N_{\rm eff} =  3.50 \pm 0.20 \, .
\end{equation}
The corresponding contours are shown in
Figure~\ref{fig:ypdhcont}. Thus, it appears
that even with the most recent reappraisals 
of the primordial abundance of $^4$He by \citet{IzoStaGus13} and \citet{Ave13},
there is better agreement (within the standard model) between
(D/H)$_{\rm p}$ and the CMB, than between (D/H)$_{\rm p}$ and \yp.

%%%%%%%%%%%%
% FIGURE 6 %
%%%%%%%%%%%%

\subsection{Deuterium and the Lepton Asymmetry}

In the past, the primordial deuterium abundance has been commonly
used as a tool for measuring the present-day universal density
of baryons (see e.g. \citealt{Ste07}), and more recently as a probe of the
effective number of neutrino families
\citep[][see also Section~\ref{sec:neffobary}]{Cyb04,NolHol11,PetCoo12}.
Here, we demonstrate that \textit{precise} measures of the primordial
deuterium abundance (in combination with the  CMB)
can also be used to estimate the neutrino degeneracy parameter, $\xi$,
which is related to the lepton asymmetry by Equation 14
from \citet{Ste12}.

\citet{Ste12} recently suggested that combined estimates for
\dhp, \yp, and a measure of \neff\ from the CMB, can provide interesting
limits on the neutrino degeneracy parameter ($\xi \le 0.079$, $2\sigma$; see also,
\citealt{SerRaf05}; \citealt{PopVas08}; and \citealt{SimSte08}). By combining
\dhp\ and \yp, this approach effectively removes the dependence on \Obary\hsq.
Using the conversion relations for \dhp\ and \yp\ (eqs.~\ref{eqn:dhptoetad}--\ref{eqn:etad}
and \ref{eqn:yptoetad}--\ref{eqn:etahe})
and the current best determination of \yp\ ($0.253\pm0.003$; \citealt{IzoStaGus13}),
in addition to the \textit{Planck}+WP+highL\footnote{We used the base
cosmology set with \neff\ and \yp\ added as free parameters
(see Section 6.4.5 of \citealt{Efs13}).} constraint on \neff\ and
the precise determination of \dhp\ reported here, we derive a $2\sigma$
upper limit on the neutrino degeneracy parameter, $|\xi|\le0.064$, based on the 
approach by \citet{Ste12}.

We propose that an equally powerful technique for estimating $\xi$ 
does \textit{not} involve removing the dependence on \Obary\hsq\ by
combining \dhp\ and \yp, as in \citet{Ste12}. Instead, one can obtain
a measure of both \Obary\hsq\ and \neff\ from the CMB, and use \textit{either}
\dhp\ or \yp\ to obtain two separate measures of $\xi$. This has the clear
advantage of decoupling \dhp\ and \yp; any systematic biases in either
of these two values could potentially bias the measure of $\xi$.
Separating \dhp\ and \yp\ also allows one to check that the two
estimates agree with one another.

Our calculation involved a Monte Carlo
technique, whereby we generated random values from the Gaussian-distributed
primordial D/H abundance measurements, whilst simultaneously drawing
random values from the (correlated) distribution between
\Obary\hsq\ and \neff\ from the \textit{Planck}+WP+highL CMB data
\citep{Efs13}\footnote{Rather than drawing values of \Obary\hsq\ and \neff\
from the appropriate distribution, we instead used the
Markov-Chain Monte Carlo chains provided by the \textit{Planck} science
team, which are available at:\\ \texttt{http://www.sciops.esa.int/wikiSI/planckpla/index.php?\\title=Cosmological\_Parameters\&instance=Planck\_Public\_PLA}}.
Using Equation~19 from \citet[][equivalent to eq.~\ref{eqn:etad} here]{Ste12},
we find $\xi_{\rm D}=+0.05\pm0.13$ for
\dhp, leading to a $2\sigma$ upper limit of
$|\xi_{\rm D}|\le0.31$.

With the technique outlined above, we have also computed the
neutrino degeneracy parameter from the current observational bound on \yp.
For this calculation, we have used the MCMC chains from the
\textit{Planck}+WP+highL CMB base cosmology with \neff\ and
\yp\ added as free parameters. In this case, the CMB distribution
was weighted by the observational bound on \yp\
(\yp\,=\,$0.253\pm0.003$; \citealt{IzoStaGus13}).
Using Equations~19--20 from 
\citet[][equivalent to eqs.~\ref{eqn:etad} and \ref{eqn:etahe} here]{Ste12},
we find $\xi_{\rm D}=+0.04\pm0.15$ for
\dhp\ and $\xi_{\rm He}=-0.010\pm0.027$ for \yp.
These values translate into
corresponding $2\sigma$ upper limits
$|\xi_{\rm D}|\le0.34$ and
$|\xi_{\rm He}|\le0.064$.
Combining these two constraints then gives $\xi=-0.008\pm0.027$,
or $|\xi|\le0.062$ ($2\sigma$).

Alternatively, if we assume that the effective number of neutrino
species is consistent with three standard model neutrinos (i.e. \neff~$\simeq3.046$),
we obtain the following BBN-only bound on the neutrino degeneracy parameter
by combining \dhp\ and \yp,
$\xi=-0.026\pm0.015$, or $|\xi|\le0.056$ ($2\sigma$).
We therefore conclude that all current estimates of the
neutrino degeneracy parameter, and hence the lepton
asymmetry, are consistent with the standard model
value, $\xi=0$.

From the above calculations, it is clear that \yp\ is the more sensitive
probe of any lepton asymmetry, as is already appreciated.
However, \dhp\ offers an additional bound on
$\xi$ that is complementary to that obtained
from \yp. We note further that, if the uncertainty in the conversion of
\dhp\ to $\eta_{\rm D}$ could be reduced by a factor of two,
the bound on $\xi_{\rm D}$ would be reduced by 15\%,
corresponding to a $1\sigma$ uncertainty on $\xi$ of $\sim0.11$
from the CMB and deuterium measurements alone.

\section{Summary and Conclusions}
\label{sec:conc}

We have reported a new precise measurement 
of the primordial abundance of deuterium
in a very metal-poor
damped Lyman\,$\alpha$ system at $z_{\rm abs} = 3.06726$
towards the QSO SDSS~J1358+6522.  
Furthermore, we have reanalyzed self-consistently
all literature systems that meet a set of strict criteria
(four cases). These criteria were imposed to identify
the systems where both accurate and precise
measures of D\,/\,H could potentially be obtained.
The new system, plus the four from the literature,
constitute the \textit{Precision Sample} of DLAs
that are best suited for a precise determination of \dhp.
Our reanalysis was performed blind (to remove human
bias), and took advantage of a new software package
that we have specifically developed for the measurement of
D\,/\,H in QSO absorption line systems. 
From the analysis of this sample,
we draw the following conclusions.

\smallskip

\noindent ~~(i) The very metal-poor DLA
([Fe/H]$\,=-2.88\pm0.03$) towards SDSS~J1358+6522
provides a strong bound on
the primordial abundance of deuterium, \dhp\ $= (2.58 \pm 0.07)\times10^{-5}$.
A weighted mean of the five systems in the Precision Sample gives
the most stringent limit on \dhp\  to date:
\dhp\ $= (2.53 \pm 0.04) \times10^{-5}$.
The corresponding baryon density is
100\,\Obary\hsq\ $=2.202 \pm 0.046$,
assuming the standard model of particle physics
with three families of neutrinos and no lepton asymmetry.
This value is in excellent agreement with 
100\,\Obary\hsq\,(CMB)\,$=2.205 \pm 0.028$
deduced from the analysis of \textit{Planck}
observations of the CMB.

\smallskip

\noindent ~~(ii) The main limitation in using \dhp\ for
cosmological parameter estimation is the conversion of
\dhp\ into the ratio of baryons-to-photons. In particular, 
modern measurements of the cross-section for the reaction \dHe\
at energies between 50--500 keV, where there currently exists
a paucity of reliable data, are highly desirable. We
estimate that a factor of two improvement in the conversion
from \dhp\  to \Obary\hsq\
would provide a measurement of the universal baryon density
from D/H as precise as that derived
from the published \textit{Planck} data.

\smallskip

\noindent ~~(iii) By combining our D/H determination
with \textit{Planck} observations of the CMB, 
we can place a tight bound
on both \Obary\hsq\ and \neff. The best-fitting parameters
so derived are 100\,\Obary\hsq=$2.23\pm0.04$ and
\neff=$3.28\pm0.28$. We therefore rule out the existence
of an additional (sterile) neutrino at $99.3$\% confidence,
provided that \neff\ and $\eta_{10}$ remained unchanged
from BBN to recombination.
The uncertainty on \neff\ could be reduced further by
a more accurate set of reaction
cross-sections for \dHe\ and, to a lesser degree, $d+d$.

\smallskip

\noindent ~~(iv) For the first time, we have combined the
\textit{Planck} CMB results with our measure of \dhp\ to place a
bound on the neutrino degeneracy parameter, $\xi_{\rm D}=+0.05\pm0.13$.
When including the current best-estimate of the
$^4$He mass fraction, \yp, derived from metal-poor H\,\textsc{ii} regions,
the combined bound on the neutrino degeneracy parameter is
$\xi=-0.008\pm0.027$, or $|\xi|<0.062$ at $2\sigma$.

\smallskip

We conclude by re-emphasizing that the most
metal-poor DLAs are potentially the best
environments where both accurate and precise
measures of the primordial abundance of deuterium
can be obtained. A combined effort to measure anew
several important nuclear reaction
cross-sections, in addition to dedicated searches
for the rare metal-poor DLAs that exhibit resolved
deuterium absorption, are the next steps necessary to pin
down further the value of \dhp. 
This approach offers a
promising and exciting avenue to test for departures
from the standard model of cosmology
and particle physics.

\section*{Acknowledgments}
We are grateful to 
the staff astronomers at the ESO VLT and Keck  
Observatory for their assistance 
with the observations.
We are indebted to Gary Steigman
for kindly communicating ahead of publication
the latest fitting formulae used in this work, and
for providing valuable comments on the manuscript.
We also thank an anonymous referee who provided
valuable comments that improved the presentation of this work.
Discussions, advice and help with various 
aspects of the work described in this 
paper were provided by Antony Lewis, Jordi Miralda-Escud{\'e},
Paolo Molaro, Ken Nollett, Pasquier Noterdaeme, John O'Meara,
Jason X. Prochaska, Signe Riemer-S{\o}renson, Donatella Romano,
and John Webb. We thank the Hawaiian
people for the opportunity to observe from Mauna Kea;
without their hospitality, this work would not have been possible.
R.~J.~C. is partially supported by NSF grant AST-1109447.
R.~A.~J. is supported by an NSF Astronomy and Astrophysics
Postdoctoral Fellowship under award AST-1102683.
M.~T.~M. thanks the Australian Research Council 
for a QEII Research Fellowship (DP0877998)
and a Discovery Project grant (DP130100568).

\appendix

\section{Model fits to literature systems}
%%%%%%%%%%%%%%%%%%%%%%%%%%%%%%%
%%%%%%%%%%%%%%%%%%%%%%%%%%%%%%%
%%%%%%%%%%%%%%%%%%%%%%%%%%%%%%%
%%%%%%%%%%%%%%%%%%%%%%%%%%%%%%%
%
%   Figures for Literature Systems
%
%

In this Appendix we show portions of
the spectra of the metal-poor DLAs from the
literature that were reanalyzed in the present work.
Further details are given in Section~\ref{sec:litsyst}.

%%%%%%%%%%%%
% FIGURE A1 %
%%%%%%%%%%%%
\begin{figure*}
  \centering
  \includegraphics[angle=0, height=21.65cm]{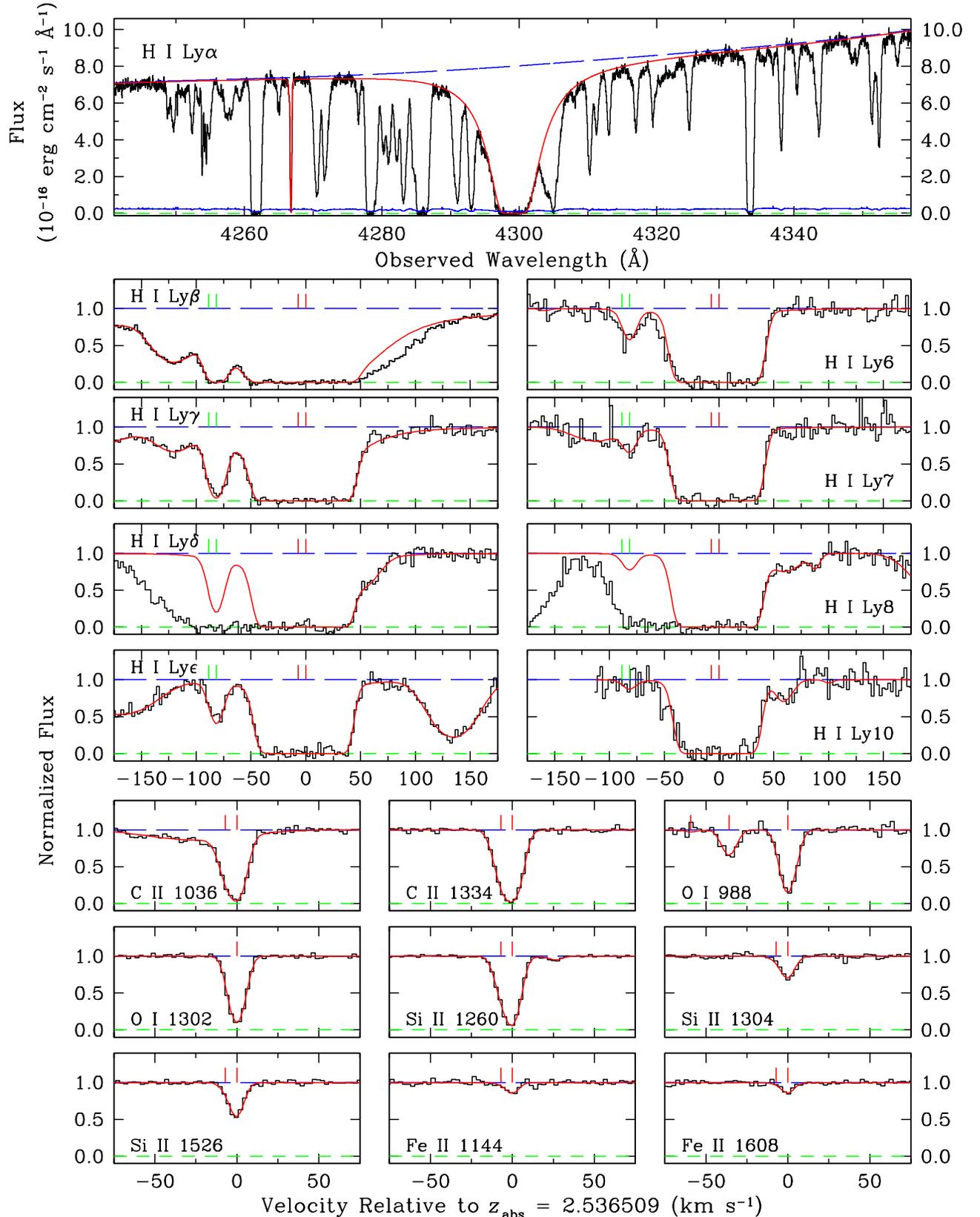}
  \caption{ 
The data (black histogram) and best-fitting model (continuous red line)
for the DLA at $z_{\rm abs}=2.53651$
toward the QSO HS\,0105$+$1619 (cf. \citealt{OMe01}).
The long-dashed blue line in the top panel marks the best-fitting
continuum level near the \Lya\ absorption line. In the remaining
panels, the data have been normalized to the continuum.
The minimum $\chi^{2}$/dof for this fit is $2088.1/3038$.
The red tick marks above the spectrum indicate the absorption
components in this DLA. The green tick marks in the Lyman
series panels show the corresponding location of \DI\ absorption.
The absorption feature near $\lambda\simeq 4267$\,\AA\ in the top panel
is \SiIII\,$\lambda\,1206.5$ at the redshift of the DLA. 
The feature near $+25$ km s$^{-1}$
in the \SiII\,$\lambda1260$ panel is \FeII\,$\lambda1260.5$.
 }
  \label{fig:HS0105p1619}
\end{figure*}

%%%%%%%%%%%%
% FIGURE A2 %
%%%%%%%%%%%%
\begin{figure*}
  \centering
  \includegraphics[angle=0, height=21.65cm]{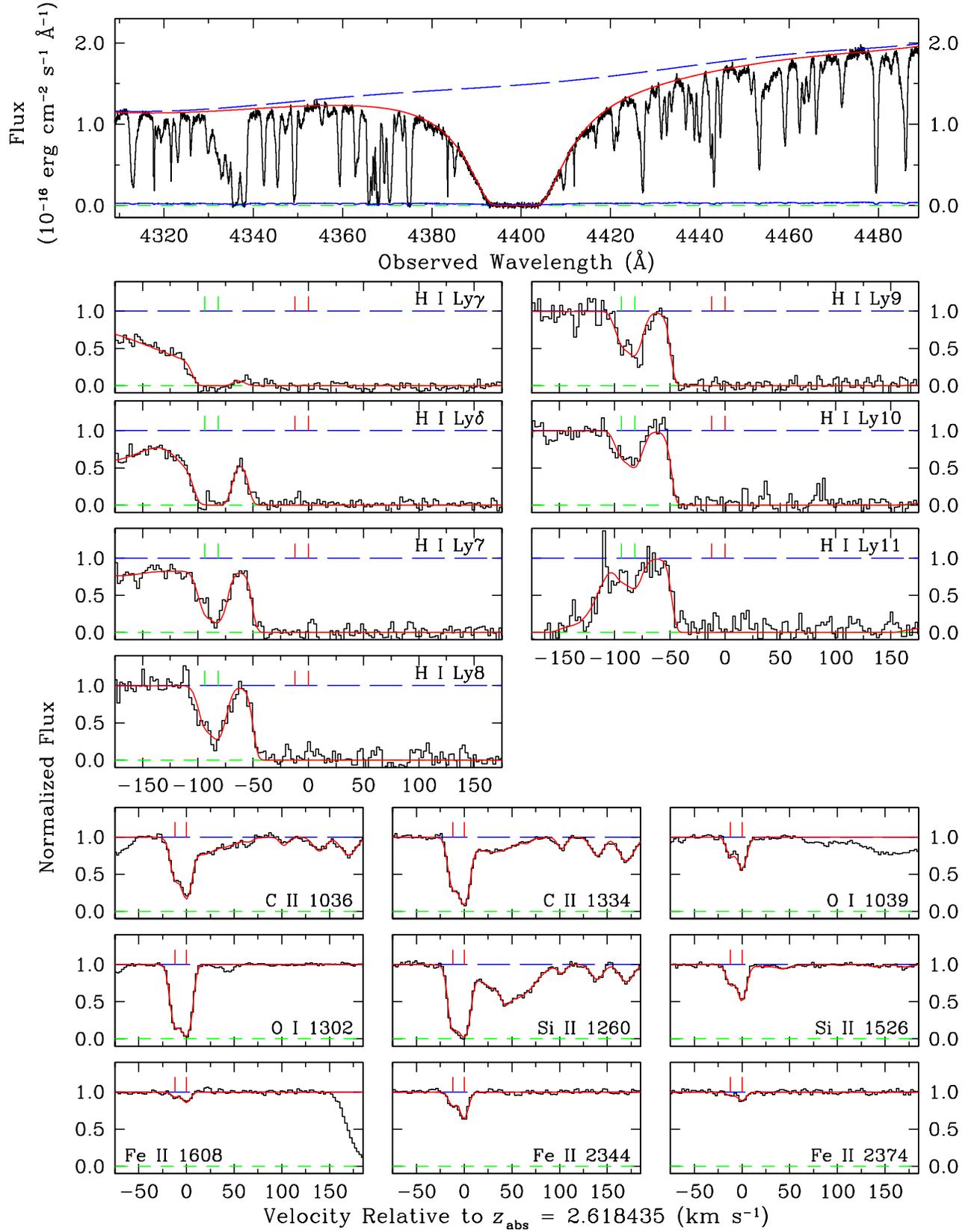}
  \caption{ 
Same as Fig.~\ref{fig:HS0105p1619} for the DLA
at $z_{\rm abs}=2.61829$ toward the QSO Q0913$+$072
(cf. \citealt{Pet08b}). The minimum $\chi^{2}$/dof for
this fit is $5539.8/4473$.
}
  \label{fig:Q0913p072}
\end{figure*}

%%%%%%%%%%%%
% FIGURE A3 %
%%%%%%%%%%%%
\begin{figure*}
  \centering
  \includegraphics[angle=0, height=21.65cm]{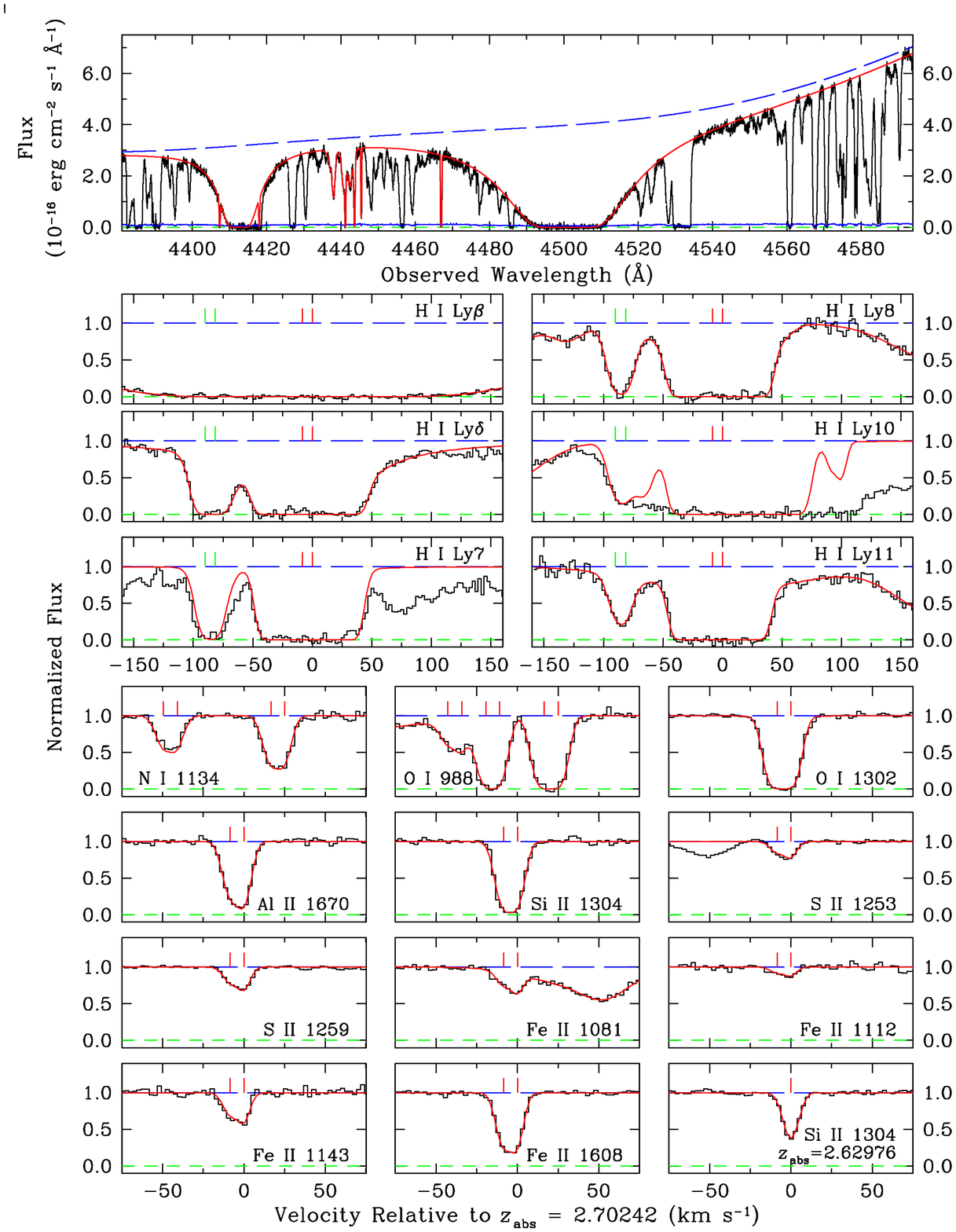}
  \caption{ 
Same as Fig.~\ref{fig:HS0105p1619} for the DLA at
$z_{\rm abs}=2.70242$ toward the QSO J1558$-$0031
(cf. \citealt{OMe06}). The DLA near $4500$\,\AA\ is the system
where D\,/\,H can be measured. The lower redshift system
(at $z_{\rm abs}=2.62976$) is included in the simultaneous
fit to the higher redshift system (see the bottom-right panel for
an example \SiII\ line profile in the low-redshift system). 
The minimum $\chi^{2}$/dof for this fit is $4754.4/4700$.
In the top panel, several absorption features from 
metal lines at $z_{\rm abs}=2.70242$ 
(\SiII\,$\lambda\lambda1190.4, 1193.3$
near 4407\,\AA\ and 4418\,\AA\ respectively,
\NI\,$\lambda\lambda1200$ and including some
additional blends near $4440$\,\AA, and \SiIII\,$\lambda1206.5$ near
$4467$\,\AA) are indicated in red.
The `extra' absorption in the trough of \HI\ Ly10 is due
to \HI\ Ly$\epsilon$ from the system at redshift $z_{\rm abs}=2.62976$.
The absorption feature at $+100$\,km\,s$^{-1}$ in this same panel
corresponds to \OI\,$\lambda 919.6$ at $z_{\rm abs}=2.70242$.
 }
  \label{fig:J1558m0031}
\end{figure*}

\end{document}